\nopagebreak  
\RequirePackage[2020-02-02]{latexrelease}
\documentclass[aps,reprint,10pt,showkeys, superscriptaddress,showpacs]{revtex4-2}
\usepackage{multirow}
\usepackage{enumitem}
\usepackage{amsfonts} 
\usepackage{amsmath}
\usepackage{amssymb}
\usepackage{graphicx}
\usepackage{subfigure}
\usepackage{color}
\usepackage{enumerate}
\usepackage[]{natbib}
\usepackage{sidecap}
\usepackage{soul}
\usepackage{cancel}
\usepackage{ulem}
\usepackage{dcolumn}%
\usepackage{bm}%

\newcommand*\xbar[1]{%
  \hbox{%
    \vbox{%
      \hrule height 0.5pt 
      \kern0.5ex
      \hbox{%
        \kern-0.1em
        \ensuremath{#1}%
        \kern-0.1em
      }%
    }%
  }%
} 

\begin{document}

\title[]{Coupling of Alfv{\'e}n and magnetosonic waves in rotating Hall magnetoplasmas}  
 \author{Amar P. Misra}%
 \homepage{Author to whom any correspondence should be addressed}
 \email{apmisra@visva-bharati.ac.in}
 \affiliation{Department of Mathematics, Siksha Bhavana, Visva-Bharati University, Santiniketan-731 235,  India} 
\author{Rupak Dey}
\email{rupakdey456@gmail.com }
\affiliation{Department of Mathematics, Siksha Bhavana, Visva-Bharati University, Santiniketan-731 235,  India} 
\author{Vinod Krishan}%
\email{vinod@iiap.res.in}
 \affiliation{Indian Institute of Astrophysics, Bangalore-560 034, India}

\date{\today}

\begin{abstract}
We study the linear theory of magnetohydrodynamic (MHD) waves, namely the Alfv{\'e}n and the fast and slow magnetosonic modes in a rotating Hall-MHD plasma with the effects of the obliqueness of the external magnetic field and the Coriolis force and show that these waves can be coupled either by the influence of the Coriolis force or the Hall effects. To this end, we derive a general form of the linear dispersion relation for these coupled modes by the combined influence of the Coriolis force and the Hall effects and analyze numerically their characteristics in three different plasma-$\beta$ regimes: $\beta\sim1$, $\beta>1$, and $\beta<1$, including some particular cases. We show that while the coupling between the Alfv{\'e}n and the fast magnetosonic modes is strong in the low-$\beta$ $(\beta\lesssim1)$ regime and the wave dispersion appears in the form of a thumb curve, in the high-$\beta~(\beta>1)$ regime, the strong coupling can occur between the Alfv{\'e}n and the slow magnetosonic modes and the dispersion appears in the form of a teardrop curve. Switching of the coupling in the  regime of $\beta\sim1$ can occur, i.e., instead of a thumb curve, a teardrop curve appears when the obliqueness of propagation and rotational angle are close to $70^\circ$ or more (but less than $90^\circ$).
 Implications of our results to solar and fusion plasmas are briefly discussed.
\end{abstract}
\maketitle
\section{Introduction}\label{intro}
In classical magnetohydrodynamic (MHD) plasmas,  the characteristic time scale of plasma oscillations is much longer than the inverse of the ion-cyclotron frequency, and electrons and ions become tied to the magnetic field, i.e., the electron and ion motions get coupled. In this case, a single fluid MHD model is applicable for describing magnetic field dynamics. However, electron and ion motions can become decoupled when the characteristic time scale is comparable to the inverse of the ion cyclotron frequency, the length-scale of magnetic field variation is akin to or smaller than the ion skin depth, or when plasma is partially ionized. In this situation,  electrons drift through ions instead of being carried along with the bulk velocity field, leading to a modification of the generalized Ohm's law by the Hall current, proportional to ${\bf J} \times {\bf B}$ force, where ${\bf J}$ and ${\bf B}$ are the current density and magnetic field respectively. Thus, the classical MHD gets modified to the Hall-MHD (HMHD). The ideal HMHD model has many applications in various astrophysical, space, and laboratory environments, e.g., flux expulsion in neutron star crusts \cite{skiathas2024combined}, formation of intensive solar flux tubes and waves in solar wind \cite{pandey2013magnetic, miteva2003surface}, propagation of whistlers in Earth's ionosphere \cite{aburjania2005generation}, fusion plasmas \cite{pandey1995hall}, dynamo mechanisms \cite{lingam2016hall}, magnetic reconnection \cite{morales2005hall} and accretion \cite{bai2017hall}, etc.
\par 
The existence of MHD waves, especially the  Alfv{\'e}n wave,  was first predicted theoretically by Alfv{\'e}n in $1942$ \cite{alfven1942existence} owing to their relevance in the Sun and astrophysical plasmas. Among the earlier researchers, Lighthill developed the standard MHD by introducing the Hall effect in 1960 \cite{lighthill1960studies} owing to their relevance in space and astrophysical plasmas. In the last few decades, many researchers have studied linear and nonlinear properties of MHD waves in Hall magnetoplasmas \cite{ruderman2020kadomtsev, almaguer1992linear, pandey2008hall}. In highly ionized plasmas, the Hall effect appears due to the inertial difference between electrons and ions. However, in partially ionized plasmas, the Hall effect may instead appear as the ions are easily decoupled (compared to electrons) from the magnetic field due to collisions with neutral atoms \cite{morales2005hall}. 
 In HMHD plasmas, Kawazura showed that the relativistic factor can influence the characteristics of the phase and group velocities of MHD waves \cite{kawazura2017modification}. Ruderman  \cite {ruderman2020kadomtsev} studied the nonlinear theory of MHD waves in anisotropic Hall plasmas by deriving a Kadomtsev-Petviashvili (KP) equation. He showed that the fast and slow magnetosonic KP solitons may become unstable due to transverse perturbations. Recently, Mahajan \textit{et al.} \cite{mahajan2024} showed how linear waves in HMHD originate a fundamental departure from the standard MHD waves, and the Hall current induces a new mode of circularly polarized waves.
\par 
Chandrasekhar and Mon \cite{chandrasekhar1953problems} first studied the rotational effects in plasma dynamics and explored the importance of the Coriolis force in the cosmic phenomena, which was later supported in subsequent studies in space and astrophysical plasmas by Lehnert \cite{lehnert1954magnetohydrodynamic}. The influence of Hall electromotive forces on the linear stability of protostellar disks was examined by Steven {\it et al.}  in rotating Hall plasmas \cite{balbus2001}. On the other hand, Rax \textit{et al.} \cite{rax2023rotating} investigated the dynamics of torsional Alfv{\'e}n waves in rotating plasmas and found two new coupling between the orbital angular momentum of the Alfv{\'e}n waves and the angular momentum of the rotating plasma.  Several authors have paid attention to studying different kinds of instabilities, including magnetogravitational instabilities and Jeans instabilities in rotating plasmas in the contexts of space and astrophysical environments \cite{turi2022magnetohydrodynamic, sharma2021modes}. In the nonlinear regime, Hager \textit{et al.} \cite{hager2023magnetosonic} studied the characteristics of magnetosonic wave propagation by deriving a Korteweg-de Vries  (KdV) equation in rotating quantum plasmas. However, they neglected the Hall effect in the linear and nonlinear analyses. Turi and Misra \cite{turi2024nonlinear} have studied the modulational instability of fast magnetosonic waves by deriving a nonlinear Schr{\"o}dinger (NLS) equation in rotating low-beta inhomogeneous magnetoplasmas without Hall current effects. The developments in the nonlinear interactions of MHD waves are also seen, e.g., in the context of solar plasmas \cite{ballester2020} and MHD wave turbulence \cite{andres2017}. However, none of the above works have focused on the fundamental MHD wave couplings by the influences of Hall effects and the Coriolis force.
\par 
 The purpose of this work is to identify the fundamental MHD modes, namely the Alfv{\'e}n and fast and slow magnetosonic modes, and how the Hall effects and the Coriolis force can contribute to the wave coupling (with switching of the couplings between the modes) of Alfv{\'e}n and magnetosonic waves, and the roles of the combined influences of these forces in rotating Hall magnetoplasmas.   
\section{Theoretical Model and Basic Equations} \label{Sec-Model}
\begin{figure} 
\centering
\includegraphics[width=3.5in,height=3.0in]{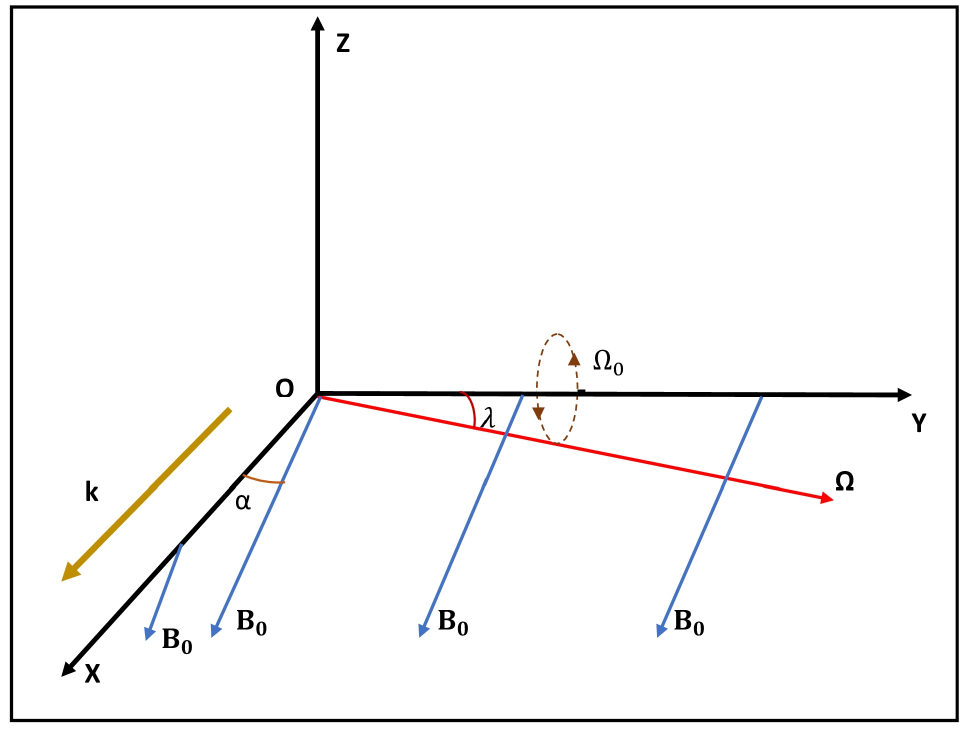}
\caption{A schematic diagram showing the geometry of the model configuration.}
\label{fig-geometry}
\end{figure}

We consider the propagation of MHD waves in a magnetized rotating Hall-MHD plasma. 
We assume that the plasma conductivity is high, i.e., the magnetic Reynolds number is larger than unity such that the magnetic field lines tend to remain frozen into the plasma and move along the fluid flow. In this situation, the term proportional to the plasma resistivity (magnetic diffusion) remains smaller than the fluid flow term (magnetic induction or advection) in the magnetic induction equation \eqref{eq-3} and so the plasma resistivity effect can be safely neglected. Also, we consider the case of high Reynolds number for which the viscous force can be neglected compared to the inertial force and that the electron and ion collision time scales are much longer than the hydrodynamic time scale of plasma oscillations for which plasma can be treated as collisionless, relevant for space plasmas. Furthermore, the length-scale of temperature variation is much larger than the fluid density variation for which the thermal conduction effect can be ignored. The plasma is supposed to be rotating about the $y$-axis with uniform angular velocity $
{\bf \Omega}_0=\left(\Omega_0\sin{\lambda},  \Omega_0\cos{\lambda}, 0\right)$,
 where $\lambda$ is the angle made by the axis of rotation with the $y$-axis. We also consider the wave propagation vector ${\bf k}$ along the $x$-axis and the uniform external  magnetic field is in the $xy$-plane, i.e., ${\bf B_0}=\left(B_0\cos{\alpha}, B_0\sin{\alpha}, 0\right)$, where $\alpha$ is the angle between ${\bf B}_0$ and ${\bf k}$. A schematic diagram for the system configuration is shown in Fig. \ref{fig-geometry}. 
 \par 
 For the description of MHD waves, we consider a two-fluid Hall-MHD model in which electron and ion fluids are treated as separate fluids. Each fluid satisfies its own mass, momentum, and energy conservation equations.
\par
The mass conservation equations for electrons and ions are
 \begin{equation}
 \frac{\partial \rho_e}{\partial t}+\nabla\cdot (\rho_e{\bf v}_e)=0, \label{eq-cont-e}
 \end{equation}
 \begin{equation}
 \frac{\partial \rho_i}{\partial t}+\nabla\cdot (\rho_i{\bf v}_i)=0, \label{eq-cont-i}
 \end{equation}
 where $\rho_j\equiv m_jn_j$ is the mass density (in which $m_j$ is the mass and $n_j$ the number density) and ${\bf v}_j$ the velocity of $j$-th species fluid with $j=e$ for electrons and $j=i$ for ions.
 \par 
 The momentum conservation equations for electrons (in the absence of the thermal pressure and collisions) reads
 \begin{equation}\label{eq-momnt-e}
\rho_e\left[\frac{\partial {\bf v}_e}{\partial t}+({\bf v}_e\cdot\nabla){\bf v}_e\right]=-en_e\left[{\bf E}+{v}_e\times {\bf B}\right]-\rho_e\left(2{\bf \Omega}_0 \times{\bf v}_e\right),
\end{equation} 
where $e$ is the elementary charge and ${\bf E}$ and ${\bf B}$ are the electromagnetic fields.
Neglecting the electron inertia, we obtain from Eq. \eqref{eq-momnt-e} the following reduced form of the generalized Ohm's law:
\begin{equation}
{\bf E}=\frac{m_e}{e}\left({\bf v}_e\times2{\bf \Omega}_0\right)-{\bf v}_e\times{\bf B}. \label{eq-ohm1}
\end{equation}
Thus, in the limit of $m_e\to 0$, Eq. \eqref{eq-ohm1} further reduces to
\begin{equation}
{\bf E}=-{\bf v}_e\times{\bf B}. \label{eq-ohm2}
\end{equation}  
\par 
Next, from the ion momentum conservation equation, we have
 \begin{equation}\label{eq-momnt-i1}
 \begin{split}
\rho_i\left[\frac{\partial {\bf v}_i}{\partial t}+({\bf v}_i\cdot\nabla){\bf v}_i\right]=&en_i\left[{\bf E}+{v}_i\times {\bf B}\right]\\
&-\nabla P_i-\rho_i\left(2{\bf \Omega}_0 \times{\bf v}_i\right),
\end{split}
\end{equation} 
where $P_i$ is the ion pressure. Using the Ohm's law \eqref{eq-ohm2} and the quaineutrality, i.e., $n_e\approx n_i=n$, Eq. \eqref{eq-momnt-i1} reduces to 
\begin{equation}\label{eq-momnt-i2}
 \rho_i\left[\frac{\partial {\bf v}_i}{\partial t}+({\bf v}_i\cdot\nabla){\bf v}_i\right]={\bf J}\times {\bf B}-\nabla P_i-\rho_i\left(2{\bf \Omega}_0 \times{\bf v}_i\right),
\end{equation} 
where ${\bf J}=en\left({\bf v}_i-{\bf v}_e\right)$ is the current density.
\par 
We require the magnetic induction equation, which can be obtained by using the Faraday's law, i.e., $\nabla\times {\bf E}=-\partial {\bf B}/\partial t$ and the Ohm's law \eqref{eq-ohm2} as
\begin{equation}
\frac{\partial {\bf B}}{\partial t}=\nabla\times \left({\bf v}_i\times {\bf B}-\frac{{\bf J}\times{\bf B}}{en}\right), \label{eq-mag-induc}
\end{equation}
where the first and second terms on the right side of Eq. \eqref{eq-mag-induc} correspond to the ion fluid flow and the Hall effects respectively.
\par 
To close the system, we also need the energy equation, i.e., the equation of state, which in the absence of loss mechanism reads
\begin{equation}
\nabla P_i=c_s^2\nabla \rho_i, \label{eq-enrg1}
\end{equation}
where $c_s$ is the ion-acoustic speed.
\par 
It may be desirable to eliminate ${\bf J}$ by using the Amp{\'e}re's law, i.e., $\nabla\times {\bf B}=\mu_0 {\bf J}$ with $\mu_0$ denoting the magnetic permeability and replace $\rho_i$ by $\rho$, ${\bf v}_i$ by ${\bf v}$, $n_i$ by $n$, and $m_i$ by $m$.  Thus, in the limit of $m_e\to0$, the required Hall-MHD can be described by the following set of fluid equations   \cite{dedner2002hyperbolic,bera2022effect}.
\begin{equation}\label{eq-1}
\frac{\partial \rho}{\partial t}+\nabla\cdot (\rho{\bf v})=0,
\end{equation}
\begin{equation}\label{eq-2}
\frac{\partial {\bf v}}{\partial t}+({\bf v}\cdot\nabla){\bf v}=-\frac{\nabla P}{\rho}+ \frac{1}{\mu_0\rho}(\nabla \times{\bf B})\times {\bf B}-2{\bf \Omega}_0 \times{\bf v},
\end{equation} 
\begin{equation}\label{eq-3}
\frac{\partial{\bf B}}{\partial t}=\nabla \times ({\bf v}\times {\bf B})-\frac{m}{e\mu_0 \rho}\nabla\times\left[(\nabla\times {\bf B})\times {\bf B}\right],
\end{equation}
\begin{equation}\label{eq-4}
\nabla P=c_s^2\nabla \rho.  
\end{equation}
In Eq. \eqref{eq-2}, we have neglected the contribution from the centrifugal force on the assumption that the rotational frequency $\Omega_0$ is small compared to the frequency of hydrodynamic oscillations.  
  \par 
 It is pertinent to express the basic equations \eqref{eq-1}-\eqref{eq-4} in dimensionless forms. So, we redefine the variables as follows:
\begin{equation}
\begin{split}
&\rho\to\rho/\rho_0, ~ {\mathbf{v}}\to {\mathbf{v} }/V_{\rm{A}}, ~{\mathbf{B}} \to {\mathbf{B}}/B_0, \\
&\left(x,y,z\right)\to\left(x,y,z\right)/\lambda_i, ~t\to \omega_{\rm{ci}}t,\\
&{\mathbf{\Omega}_0} \to {\mathbf{ \Omega}_0}/\omega_{\rm{ci}}, ~P \to P/{V_{\rm{A}}^2\rho_0},\\
\end{split}
\end{equation}
where $\rho_0=n_0m$ is the unperturbed value of $\rho$ with $n_0$ denoting the unperturbed number density  of electrons and ions, $V_{\rm{A}}=B_0/\sqrt{\mu_0\rho_0}$ is the Alfv{\'e}n speed,  $\omega_{\rm{ci}}~(\equiv eB_0/m_i)$ is the ion-cyclotron frequency, and $\lambda_i= V_{\rm{A}}/\omega_{\rm{ci}}$ is the ion skin depth (or inertial length).  Thus, Eqs. \eqref{eq-1}-\eqref{eq-4} reduce to   
\begin{equation}\label{eq-n1}
\frac{\partial \rho}{\partial t}+\nabla\cdot (\rho{\mathbf{v}})=0,
\end{equation}
\begin{equation}\label{eq-n2}
\frac{\partial {\mathbf{v}}}{\partial t}+({\mathbf{v} }\cdot\nabla){\mathbf{v} }=-\frac{\nabla P}{\rho}+ \frac{1}{\rho}(\nabla \times{\mathbf{B}})\times {\mathbf{B}}-2{\mathbf{\Omega}_0 } \times{\mathbf{v}},
\end{equation} 
\begin{equation}\label{eq-n3}
\frac{\partial{\mathbf{B}}}{\partial t}=\nabla \times ({\mathbf{v}}\times {\mathbf{B}})-\frac{1}{\rho}\nabla\times\left[(\nabla\times {\mathbf{B}})\times {\mathbf{B}}\right],
\end{equation}
\begin{equation}\label{eq-n5}
\nabla P=\tilde{c}_{\rm{s}}^2\nabla \rho,
\end{equation}
where $\tilde{c}_{\rm{s}}^2= c_{\rm s}^2/V_{\rm A}^2$, and we define it as the plasma beta, i.e., $\beta=\tilde{c}_{\rm{s}}^2$. 

\section{Dispersion relation}\label{Sec-linear}
We study the excitation of fundamental MHD wave eigenmodes and their possible coupling by the influences of the Coriolis force in the fluid motion and the Hall resistance in the magnetic induction. Specifically, we consider three different plasma-$\beta$ regimes: $\beta\sim1$, $\beta>1$, and $\beta<1$, which, respectively, correspond to the cases of ${c}_{\rm s}\sim V_{\rm A}$, ${c}_{\rm s}> V_{\rm A}$, and ${c}_{\rm s}< V_{\rm A}$, to study the acoustic or magnetic characters (or both) of MHD wave perturbations in the small-amplitude limit for which the linear theory is valid. To this end, we linearize Eqs. \eqref{eq-n1}-\eqref{eq-n5} about the equilibrium state of dependent physical variables and split up the physical quantities into their equilibrium  ($0$, or with suffix $0$) and perturbation (with suffix $1$) parts according to $\rho=1+\rho_1, ~{\mathbf{v}}=0+{\mathbf{v}_1},  ~{\mathbf{B}}=\hat{\mathbf{B}}_0+\mathbf{B}_1$, and $P=P_0+P_1$, and assume all the perturbations to vary as plane waves with the wave frequency $\omega$ (normalized by $\omega_{ci}$) and the wave vector ${\bf k}$ (normalized by $\lambda_i$) in the form $\exp\left[i\left({\bf k}\cdot {\bf r}-\omega t\right)\right]$. Thus, we obtain from Eqs. \eqref{eq-n1}-\eqref{eq-n5} the following linearized equations for the perturbations.
\begin{equation}\label{eq-p1}
\omega \rho_1-{\bf k}\cdot{\bf v}_1=0,
\end{equation}
\begin{equation}\label{eq-p2}
\omega{\bf v}_1={\bf k}P_1+ \left[\left({\hat{ \bf B}}_0\cdot{\bf B}_1\right){\bf k}-\left({\hat{ \bf B}}_0\cdot {\bf k}\right){\bf B}_1\right]-2i{\bf \Omega}_0 \times{\bf v}_1,
\end{equation} 
\begin{equation}\label{eq-p3}
\omega{\bf B}_1+\left[\left({\bf k}\cdot{\hat{\bf B}}_0\right){\bf v}_1-\left({\bf k}\cdot {\bf v}_1\right){\hat{\bf B}}_0\right]=
 i\left({\hat{\bf B}}_0\cdot {\bf k} \right)\left({\bf k}\times {\bf B}_1\right),
\end{equation}
\begin{equation}\label{eq-p5}
P_1=\tilde{c}_s^2\rho_1,
\end{equation}
where $\hat{\bf B}_0=\left(\cos \alpha, ~\sin \alpha, ~0\right)$.  We note that the ratio of the last two terms on the right side of Eq. \eqref{eq-p2} scales as (in dimensions)
\begin{equation}
\begin{split}
 \frac{B_0 B_1 k}{\rho_0\mu_0} \frac{1}{2\Omega_0v_1}&\sim \frac{kB_0 }{\sqrt{\rho_0\mu_0}}\frac{B_1}{v_1\sqrt{\rho_0\mu_0}}\frac{1}{2\Omega_0}\\
 &\sim \frac{k V_A}{2\Omega_0} \frac{B_1}{B_0}\frac{V_A}{v_1}\sim \frac{\omega_A}{2\Omega_0},
 \end{split}
\end{equation}   
where $\omega_A=kV_A$ is the Alfv{\'e}n frequency and $B_1/B_0\sim v_1/V_A$. Thus, for Alfv{\'e}n waves, the Coriolis force dominates over the ${\bf J}\times {\bf B}$-force for $\omega_A<2\Omega_0$.
 Assuming, for simplicity, the wave propagation vector along the $x$-axis, i.e., ${\bf k}=(k, 0,0)$ so that ${\bf k}\cdot \hat{\bf B}_0=k\cos{\alpha}$ and looking for nonzero solutions of the perturbations, we obtain from Eqs. \eqref{eq-p1}-\eqref{eq-p5} the following linear dispersion relation.
\begin{widetext}
\begin{equation}\label{eq-disp-eta0}
\begin{split}
&\left(\omega^2-k^2\cos^2{\alpha}\right)\left[\omega^4-(1+\tilde{c}_s^2)k^2\omega^2+\tilde{c}_s^2k^4\cos^2{\alpha}\right]-4\Omega_0^2 \omega^2\left[\omega^2-\tilde{c}_s^2 k^2\sin^2{\lambda}- k^2\cos^2{\left(\alpha+\lambda\right)}\right] \\
& -\omega^2k^4 \cos^2{\alpha} \left(\omega^2-\tilde{c}_s^2k^2\right)+4\Omega_0k^4\cos^2{\alpha}\left[\omega^2 \sin{\left(\alpha+\lambda\right)}-\tilde{c}_s^2k^2\cos{\alpha}\sin{\lambda}\right]+4\Omega_0^2k^4\cos^2{\alpha}\left(\omega^2-\tilde{c}_s^2k^2\sin^2{\lambda}\right)=0.
\end{split}
\end{equation}
\end{widetext}
Equation \eqref{eq-disp-eta0} is the general form of dispersion relation for MHD waves in an electron-ion magnetoplasma with the influences of the Coriolis force and Hall resistance. It generalizes some previous works (e.g., Refs. \cite{ruderman2020kadomtsev,turi2022magnetohydrodynamic}) where the combined influences of the Coriolis force and the Hall resistance on the wave modes have not been discussed. By considering the wave propagation parallel to the magnetic field (i.e., $\alpha=0$) and ignoring the Hall effect, one can recover the same dispersion relation as in Ref. \cite{turi2022magnetohydrodynamic} after setting $\theta=0, ~\nu_c=0$, and $\omega_J=0$ therein.  In the absence of the Coriolis force and with a substitution of $\omega^2$ by $\omega k\cos\alpha$ in the Hall contributed term, the dispersion equation \eqref{eq-disp-eta0} reduces to Eq. (8) of Ref. \cite{ruderman2020kadomtsev}. 
\par 
 Several terms appearing in Eq. \eqref{eq-disp-eta0} correspond to different physical sources. On the left side, the second term ($\propto 4\Omega_0^2\omega^2$) appears due to the effects of the Coriolis force, the third term ($\propto k^4\cos^2\alpha$) appears due to the Hall effect, and the fourth ($\propto 4\Omega_0 k^4\cos^2\alpha$) and fifth ($\propto 4\Omega_0^2 k^4\cos^2\alpha$) terms appear due to the combined influences of the Coriolis force and Hall resistance. Furthermore, in the first term, while the first factor  corresponds to the shear Alfv{\'e}n wave, the second factor gives fast and slow magnetosonic modes. This can be verified by disregarding the Coriolis force and Hall effects in Eq. \eqref{eq-disp-eta0}. Thus, the coupling between the Alfv{\'e}n and magnetosonic modes can occur by the influence of either the Coriolis force, or the Hall effect, or both of them. When these effects are absent, Eq. \eqref{eq-disp-eta0} gives three decoupled MHD mods: the shear Alfv{\'e}n wave and the fast and slow magnetosonic modes, to be discussed shortly in Sec. \ref{sec-sub-basic}.   
Before we study the coupling of the MHD waves and their characteristics in a general situation, it may be helpful to discuss about some particular cases that correspond to some  known results in the literature and to demonstrate relative influences of the Coriolis force and the Hall effect on the MHD modes.   
\subsection{Wave motion without the Coriolis force and Hall effects}\label{sec-sub-basic}
We consider the MHD wave propagation in a non-rotating plasma with a frequency much lower than the ion-cyclotron frequency. In this case, the Hall effect can be neglected and there will be no effect of the  Coriolis force. As a result, the dispersion relation \eqref{eq-disp-eta0} reduces to that for decoupled Alfv{\'e}n and magnetosonic modes, i.e., 
\begin{equation}\label{eq-disp-none}
\left(\omega^2-k^2\cos^2{\alpha}\right)
\left[\omega^4-\left(1+\tilde{c}_s^2\right)k^2\omega^2+\tilde{c}_s^2k^4\cos^2{\alpha}\right]=0.
\end{equation}
The first factor of Eq. \eqref{eq-disp-none}, when equated to zero, gives the non-dispersive shear Alfv{\'e}n mode \cite{murawski1992alfven}, given by,
\begin{equation}\label{eq-disp-none-Alf}
\omega=k\cos \alpha.
\end{equation}  
Such transverse waves, driven by the magnetic tension, can not propagate perpendicular to the static magnetic field but with an angle satisfying $0\leq\alpha<\pi/2$. Since there is no density or pressure fluctuations associated with the  wave, the wave energy flows along the magnetic field lines at the Alfv{\'e}n  speed $V_{\rm A}$, which in the dimensionless form gives $\omega/k=1$ at $\alpha=0$.  On the other hand, equating the second factor of Eq. \eqref{eq-disp-none} to zero gives the following dispersion relations for the fast (with suffix `F') and slow (with suffix `S') magnetosonic modes propagating obliquely ($0<\alpha<\pi/2$) to the magnetic field \cite{murawski1992alfven}.
\begin{equation}\label{eq-a}
\omega_{F,S}^2=\frac{k^2}{2}\left[\left(1+\tilde{c}_s^2\right)\pm\sqrt{\left(1+\tilde{c}_s^2\right)^2-4\tilde{c}_s^2\cos^2\alpha}\right],
\end{equation}
where the plus and minus signs, respectively, correspond to the fast and slow magnetosonic modes with frequencies $\omega_F$ and $\omega_S$. In contrast to the Alfv{\'e}n wave, the magnetosonic modes are driven by the magnetic tension and the pressure gradient forces.
In particular, for  $\alpha=\pi/2$, i.e., when ${\bf \hat B}_0 \perp {\bf k}$, the fast mode emerges as the classical magnetosonic wave: $\omega_F/k=\sqrt{1+\tilde{c}_s^2}$ while the slow mode disappears. Also, for  $\alpha=0$, i.e., for ${\bf \hat B}_0 \parallel {\bf k}$, the fast mode turns out to be an acoustic mode: $\omega_F/k=\tilde{c}_s$ and the slow mode becomes the Alfv{\'e}nic mode:  $\omega_S/k=1$ (i.e., $\omega_S=kV_{\rm A}$ in dimensional form), or vice versa, depending on the plasma beta, $\beta~(\sim \tilde {c}_s^2)>1$, or $\beta<1$. On the other hand, for the propagation of MHD waves obliquely to the magnetic field, i.e.,   $0<\alpha<\pi/2$, Eq. \eqref{eq-a} yields for the fast magnetosonic wave the frequency: (i) $\omega_F\approx \tilde {c}_sk$ in the limit of  $\beta \gg 1$, i.e., the fast mode becomes acoustic in character (longitudinal fluid motion) and (ii) $\omega_F\approx k$ in the limit  of $\beta \ll 1$, i.e., the fast mode becomes magnetic in character (fluid motion transverse to the magnetic field).     Thus, the fast mode (when the fluid and magnetic pressure variations are in phase) is basically an acoustic wave in the high-$\beta$ regimes such as those in the solar convection zone,  photosphere, and lower chromosphere. However, in the low-$\beta$ regimes  (e.g., solar corona and upper chromosphere), it becomes more like an Alfv{\'e}n wave \cite{murawski1992alfven}. On the other hand, Eq. \eqref{eq-a} gives for slow magnetosonic modes (when the fluid and magnetic pressure variations are out of phase) propagating obliquely to the magnetic field $(0<\alpha<\pi/2)$  the frequency,   $\omega\approx  k\tilde{c}_s \cos{\alpha}$ in the limit of $\beta\ll1$ and  the frequency  $\omega\approx k\cos{\alpha}$ in the limit of $\beta\gg1$. It also follows that,  in contrast to the fast mode and in the low-$\beta$ regime (e.g., solar corona), the slow magnetosonic mode is more acoustic than magnetic. Thus, it may be predicted that if there is a possibility of coupling between the Alv{\'e}n and magnetosonic modes by the influence of either the Coriolis force or the Hall effect, or both, it may be likely that in the low-$\beta$ regime, this coupling can occur between the fast magnetosonic mode (more acoustic) and the Alv{\'e}n mode (magnetic), and in the high-$\beta$ regime, the same can be between the slow magnetosonic (more acoustic) and Alv{\'e}n  (magnetic) modes. We will justify these assertions and give a clearer picture about the couplings in Secs. \ref{sec-sub-coriolis}-\ref{sec-sub-co-ha}.
\subsection{Wave motion with the Coriolis force but without the Hall effect}\label{sec-sub-coriolis}
We consider the propagation of MHD waves with the wave frequency much lower than the ion-cyclotron frequency as in Sec. \ref{sec-sub-basic} but in rotating magnetoplasmas. In this case, the Hall effect can be neglected, and the dispersion equation \eqref{eq-disp-eta0} reduces to 
 \begin{equation}\label{eq-disp-Cor}
\begin{split}
&\left(\omega^2-k^2\cos^2{\alpha}\right)\left[\omega^4-(1+\tilde{c}_s^2)k^2\omega^2+\tilde{c}_s^2k^4\cos^2{\alpha}\right]\\
&-4\Omega_0^2 \omega^2\left[\omega^2-\tilde{c}_s^2 k^2\sin^2{\lambda}- k^2\cos^2{\left(\alpha+\lambda\right)}\right]=0.
\end{split}
\end{equation}
It is evident that not only are Alfv{\'e}n and magnetosonic waves coupled but also modified by the influence of the Coriolis force (the term proportional to $\Omega_0^2$). To reveal the effects of this force on the fundamental modes, we first consider the magnetosonic mode at an angle $\alpha=\pi/2$ (at which the Alfv{\'e}n mode disappears). The  dispersion equation \eqref{eq-disp-Cor} then reduces to
\begin{equation}\label{eq-disp-Cor-alphapi}
\omega^4-\left[\left(1+\tilde{c}_s^2\right)k^2+4\Omega_0^2\right]\omega^2+4\Omega_0^2k^2\sin^2{\lambda}\left(1+\tilde{c}_s^2\right)=0,
\end{equation}
from which the frequencies for the fast and slow modes are given by  
\begin{equation} \label{eq-FS-C}
\omega_{F,S}^2=\frac{1}{2}\left[\Lambda^2\pm\sqrt{\Lambda^4-16\Omega_0^2\left(1+\tilde{c}_s^2\right)k^2\sin^2{\lambda}}\right].
\end{equation}
Here, $\Lambda^2=\left(1+\tilde{c}_s^2\right)k^2+4\Omega_0^2$, and the plus (minus) sign before the radical sign stands for the fast (slow) magnetosonic mode. The cut-off frequencies (at $k=0$) for the fast and slow modes, respectively, are $2\Omega_0$ and $0$. In particular, if $\lambda=\pi/2$, i.e., the axis of rotation coincides with the propagation vector, the fast and slow modes become decoupled with frequencies,  $\omega_F=k\sqrt{1+\tilde{c}_s^2}$ and $\omega_S= 2\Omega_0 $, i.e., while the former appears as a non-dispersive magnetosonic mode, the latter emerges as a constant oscillation with a frequency twice the plasma rotational frequency.  On the other hand, when $\lambda=0$, i.e., the axis of rotation is perpendicular to the propagation direction, the fast mode emerges as a dispersive magnetosonic mode with a frequency, $\omega=\sqrt{\left(1+\tilde{c}_s^2\right)k^2+4\Omega_0^2}$, which gets modified by the rotational frequency and agrees with Eq. (15) of Ref. \cite{turi2022magnetohydrodynamic} if one disregards the effects of the cosmic pressure and gravitational force therein. However, the slow mode disappears. The cut-off frequency for the fast magnetosonic mode is $2\Omega_0$, which is the same as obtained from Eq. \eqref{eq-FS-C}, i.e., independent of the angle of rotation.  
\par Next, we consider the propagation parallel to the magnetic field $(\alpha=0)$ and study the effects of the Coriois force on the wave dispersion. In this case, Eq. \eqref{eq-disp-Cor} reduces to
\begin{equation}\label{eq-disp-Cor-alpha0}
\begin{split}
&\left(\omega^2-k^2\right)\Big\lbrace\omega^4-(1+\tilde{c}_s^2)k^2\omega^2+\tilde{c}_s^2k^4\Big\rbrace-4\Omega_0^2 \omega^2\\
&\times \left(\omega^2-\tilde{c}_s^2 k^2\sin^2{\lambda}- k^2\cos^2{\lambda}\right)=0.
\end{split}
\end{equation}
From Eq. \eqref{eq-disp-Cor-alpha0}, a novel feature is that even in the case of MHD wave propagation parallel to the static magnetic field, not only are both the magnetic and acoustic characters of the magnetosonic modes retained, the modes also get coupled to the Alfv{\'e}n mode by the influence of the Coriolis force. The reason why the magnetic character of the magnetosonic mode persists even at $\alpha=0$ may be both the Lorentz and the Coriolis forces are similar in that both are proportional to the particle velocity and act perpendicular to it \cite{opat1990coriolis}.  In the absence of the Coriolis force, both the Alfv{\'e}n and magnetosonic modes get decoupled, giving the fast mode to appear as the Alfv{\'e}n mode $\omega=k$ and the slow mode as the acoustic mode with $\omega=\tilde{c}_s k$.   
In particular, for $\lambda=\pi/2$ (i.e., when the axis of rotation is along the propagation vector),   Eq. \eqref{eq-disp-Cor-alpha0} gives the magnetosonic wave as an acoustic mode with $\omega=\tilde{c}_s k$ and the fast and slow Alfv{\'e}n waves \cite{turi2022magnetohydrodynamic}, given by, 
\begin{equation}
\omega=\pm\Omega_0+\sqrt{\Omega_0^2+k^2}.
\end{equation}
It shows that the Alfv{\'e}n waves propagating with a frequency lower than the ion-cyclotron frequency in a rotating magnetoplasma become dispersive due to the Coriolis force. The cut-off frequency for the slow Alfv{\'e}n  mode is zero but it is shifted by $2\Omega_0$ for the fast mode.  However, at $\lambda=0$  (i.e., when the axis of rotation is perpendicular to  the wave vector), we obtain from  Eq. \eqref{eq-disp-Cor-alpha0} a non-dispersive Alfv{\'e}n mode, $\omega=k$, and dispersive fast and slow magnetosonic modes, given by,
\begin{equation}
\omega_{F,S}^2=\frac{1}{2}\left[\Lambda^2 
\pm\sqrt{\Lambda^4-4\tilde{c}_s^2k^4}\right],
\end{equation}
with the cut-off frequencies, $2\Omega_0$ and $0$, respectively.
\par 
In what follows, we consider a general situation in which $0<\alpha,~\lambda<\pi/2$, and obtain from Eq. \eqref{eq-disp-Cor} approximate dispersion relations for both the Alfv{\'e}n and magnetosonic waves. For the Alfv{\'e}n wave, we have
\begin{equation}\label{eq-disp-Cor-Alf}
\omega^2=d_1+k^2 \cos^2\alpha, 
\end{equation}
where $d_1$ is the correction term, obtained by replacing $\omega$ by $k\cos\alpha$ in the term proportional to $\Omega_0^2$ and divided by the factor associated with the magnetosonic wave, given by,
\begin{equation}
d_1\sin^2\alpha=4\Omega_0^2\left[\tilde c_s^2\sin^2\lambda- \cos^2\alpha+\cos^2\left(\alpha+\lambda\right)\right].
\end{equation}
Typically, the approximate dispersion relation \eqref{eq-disp-Cor-Alf} describes an Alfv{\'e}n mode that is hybridized by the coupling effect of the magnetosonic mode in the presence of the Coriolis force.  Furthermore, the dispersion relation is in the form of a Langmuir wave in classical plasmas. So, it can correspond to a fundamental MHD mode, whose nonlinear evolution as Alfve{\'n}ic wave envelopes can be described by a nonlinear Schr{\"o}dinger (NLS)-like equation \cite{misra2009,chatterjee2018}. 
 From Eq. \eqref{eq-disp-Cor-Alf}, it is also evident that the hybridized Alfv{\'e}n mode can propagate only for the wave number exceeding a critical value, i.e.,  $k>k_c$, where the critical value $k_c$ is given by
\begin{equation}
k_c=4\Omega_0\sqrt{ \cos^2{\alpha}-\cos^2\left(\alpha+\lambda\right)-\tilde c_s^2\sin^2{\lambda}}/\sin{2\alpha}, 
\end{equation} 
 provided $\tilde c_s^2\leq \sin{\left(2\alpha+\lambda\right)}/\sin\lambda$, or $\lambda\leq\lambda_c$ with $\lambda_c\equiv\tan^{-1}\left[ \left(1-\sin2\alpha\right)/\cos2\alpha\right]$ denoting the critical value of $\lambda$. In particular, in the long-wavelength limit,  $k\ll 1$ with $d_1>0$, Eq. \eqref{eq-disp-Cor-Alf} reduces to
\begin{equation}
\omega\approx\sqrt{d_1}+\frac12\left(\frac{\cos^2 {\alpha}}{\sqrt{d_1}}\right)k^2. 
\end{equation}
  Thus, low-frequency, long-wavelength Alfv{\'e}n waves have second order dispersion, implying that its nonlinear evolution as Alfv{\'e}nic solitary waves can not be precisely described by the Korteweg de-Vries (KdV)-like equations but NLS-like equations.
\par 
Similarly, looking for an approximate dispersion relation for the magnetosonic wave (hybridized by the coupling effect of the Alfv{\'e}n wave), we obtain from Eq. \eqref{eq-disp-Cor} the following relation. 
\begin{equation}\label{eq-disp-Cor-mag}
\omega_{F,S}^2=\frac{k}{2}\left[\left(1+\tilde c_s^2\right)k+\sqrt{b_{1\pm}+k^2\lbrace 2a^2_\pm -\left(1+\tilde c_s^2\right)\rbrace^2}\right],
\end{equation}
 where the plus and minus sign correspond to the fast and slow modes with the phase velocities, $a_{\pm}\equiv \omega_{F,S}/k$, given by Eq. \eqref{eq-a}, and $b_{1\pm}$ is given  by 
\begin{equation}
b_{1\pm}=\frac{16\Omega_0^2 a_{\pm}^2\left[ a_{\pm}^2-\tilde c_s^2\sin^2\lambda-\cos^2{\left(\alpha+\lambda\right)}\right]}{a_{\pm}^2-\cos^2\alpha}.
\end{equation}
In the long-wavelength limit  $k\ll 1$, Eq. \eqref{eq-disp-Cor-mag} reduces to
\begin{equation} \label{eq-magne-Corio}
\omega_{F,S}\approx b_{1\pm}^{1/4}K+\frac{\left(1+\tilde c_s^2\right)}{b_{1\pm}^{1/4}}K^3+\frac{\lbrace 2a_{\pm}^2-\left(1+\tilde c_s^2\right)\rbrace^2}{b_{1\pm}^{3/4}}K^5,
\end{equation}
where $K=\sqrt{k/2}$. Equation \eqref{eq-magne-Corio} discerns that low-frequency long-wavelength magnetosonic waves have a cubic order dispersion or higher. So, the corresponding weakly nonlinear evolution of magnetosonic solitary waves can be governed by a KdV-like equation with or without higher-order (than cubic) dispersion. 
\par From the results obtained in this section, we may conclude that the Alfv{\'e}n and magnetosonic waves propagating in a rotating magnetoplasma get coupled by the influence of the Coriolis force. The Coriolis force plays an important role in retaining the magnetic properties of the magnetosonic mode even when the wave propagation is parallel to the magnetic field. This is in contrast to non-rotating plasmas where the magnetosonic modes emerge as acoustic-like modes \cite{andres2017}. The reason may be that the Coriolis and Lorentz forces have mathematical similarities, i.e., they are proportional and perpendicular to the particle velocity. The Alfv{\'e}n and magnetosonic waves also become dispersive due to the Coriolis force effect with even $(\geq2)$ and odd $(\geq3)$ orders of dispersion, implying that their  nonlinear evolution as solitary waves can be described by NLS- and KdV-like equations, respectively, and the nonlinear coupling of these waves (to be governed by a coupled KdV- and NLS-like equations) could be more pronounced in the presence of the Coriolis force.  

\subsection{Wave motion with the Hall effect but without the Coriolis force}\label{sec-sub-hall} 
In this section, we consider the MHD wave propagation in a non-rotating plasma, and assume that the characteristic scale length $(k^{-1})$ is comparable to or smaller than the collisionless ion skin depth and the characteristic time scale of hydrodynamic oscillations is comparable to the ion gyroperiod such that the Hall effect may no longer be negligible. So, disregarding the terms involving $\Omega_0$ in Eq. \eqref{eq-disp-eta0}, we obtain \cite{ruderman2020kadomtsev, almaguer1992linear}
\begin{equation}\label{eq-disp-Hall}
\begin{split}
&\left(\omega^2-k^2\cos^2{\alpha}\right)\left[\omega^4-(1+\tilde{c}_s^2)k^2\omega^2+\tilde{c}_s^2k^4\cos^2{\alpha}\right] \\
& -\omega^2 k^4 \cos^2{\alpha} \left(\omega^2-\tilde{c}_s^2k^2\right)=0.
\end{split}
\end{equation}
We note that the third term on the left side of Eq. \eqref{eq-disp-Hall} appears due to the Hall effect, giving the  Alfv{\'e}n and magnetosonic modes coupled, and the coupling persists until the wave propagation direction remains oblique to the magnetic field, i.e., $0\leq\alpha<\pi/2$. 
The dispersion equation \eqref{eq-disp-Hall} agrees with Eq. (8) of Ref. \cite{ruderman2020kadomtsev} when one considers the appropriate normalizations for the physical quantities and approximates the factor $\omega^2$ by $\omega k\cos\alpha$ in the leading factor of the term associated with the Hall effect (the last term on the left side). Such an approximation may be valid when the Hall contribution remains small compared to the other effects and the characteristic length scale of wave excitation is much larger than the characteristic length of wave dispersion \cite{ruderman2020kadomtsev}. In this situation, the Hall contributed term may be considered a correction to the dispersion relation. We, however, do not make any approximation at this stage; rather, consider some particular cases.   
For example, we note that when $\alpha=\pi/2$, i.e., the wave propagation is perpendicular to the magnetic field, the Hall contribution disappears and only the magnetosonic mode emerges with a frequency, given by, $\omega^2=(1+\tilde{c}_s^2)k^2$, which means that the Hall contribution can be effective for Alfv{\'e}n and magnetosonic modes when the propagation angle remains within the interval:  $0\leq\alpha<\pi/2$.   On the other hand, for $\alpha=0$, i.e., when the wave propagation is along the magnetic field, Eq. \eqref{eq-disp-Hall} gives the magnetosonic modes to emerge as the non-dispersive acoustic mode with frequency $\omega=k\tilde c_s$ and  the 
Alfv{\'e}n mode with the following dispersion relation:
\begin{equation}\label{eq-disp-Hall-alpha0}
\omega=\frac{k}{2}\left(k+\sqrt{4+k^2}\right),
\end{equation}
in which the higher-order dispersion than the first-order (See the first term and the second term under the radical sign on the right side)  appears due to the Hall effect. 
 \par 
 Next, we consider the case of oblique propagation of MHD waves with $0<\alpha<\pi/2$ and obtain approximate dispersion relations for both Alfv{\'e}n and magnetosonic modes by the same way as in Sec. \ref{sec-sub-coriolis}. For the Alfv{\'e}n waves, modified by the coupling effect of the magnetosonic wave, we obtain the following dispersion relation.
\begin{equation}\label{eq-disp-Hall-Alf}
\omega^2=k^2 \cos^2\alpha+d_2k^4,
\end{equation}
where $d_2=\left(\tilde c_s^2-\cos^2\alpha\right)\rm{cot}^2\alpha$, which is positive in a high-beta regime $(\beta\sim\tilde {c}^2_s>1)$ and can be negative in a low-beta regime  $(\beta<1)$ depending on the angle $\alpha$.
In the long-wavelength limit, $k\ll1 $, Eq. \eqref{eq-disp-Hall-Alf} reduces to \cite{ruderman2020kadomtsev},
\begin{equation}\label{eq-disp-Hall-Alf_approx}
\omega=k \cos\alpha+\frac{1}{2}d_2k^3\sec\alpha.
\end{equation}
This hybridized low-frequency Alfv{\'e}n mode has a cubic (or higher) order dispersion, implying that it would be meaningful to describe the nonlinear evolution of small-amplitude  Alfv{\'e}n waves by a KdV-like equation.
By the same way, an approximate dispersion relation for the magnetosonic wave  can be obtained from Eq. \eqref{eq-disp-Hall}  as
\begin{equation}\label{eq-disp-Hall-mag}
\omega_{F,S}^2=\frac{k^2}{2}\left[\left(1+\tilde c_s^2\right)+\sqrt{\lbrace 2a^2_\pm -\left(1+\tilde c_s^2\right)\rbrace^2+b_{2\pm}k^2}\right],
\end{equation}
where $a_{\pm}$ is defined by Eq. \eqref{eq-a} and $b_{2\pm}$ (where the plus sign is for the fast mode and minus sign for the slow mode) is given by
\begin{equation}\label{exp-b2}
b_{2\pm}=\frac{4a_{\pm}^2\left(a_{\pm}^2-\tilde c_s^2\right)\cos^2\alpha}{\left(a_{\pm}^2-\cos^2\alpha\right)}.
\end{equation}
In the long-wavelength limit, $k\ll1$, Eq. \eqref{eq-disp-Hall-mag} reduces to \cite{ruderman2020kadomtsev} 
\begin{equation}\label{eq-disp-Hall-mag_approx}
\omega=a_{\pm}k+b_{\pm}k^3,
\end{equation}
where $a_{-}<a_{+}$ and $b_{\pm}$ are given  by 
\begin{equation}\label{exp-b}
b_\pm=\frac{a_{\pm}\left(a_{\pm}^2-\tilde c_s^2\right)\cos^2\alpha}{2\left(a_{\pm}^2-\cos^2\alpha\right)\lbrace 2a^2_\pm -\left(1+\tilde c_s^2\right)\rbrace}
\end{equation}
with $b_{+}>0$ and $b_{-}<0$. Like the Alfv{\'e}n wave, the low-frequency long-wavelength magnetosonic mode has also a cubic or higher-order dispersion for which the KdV theory may be applicable for the evolution of weakly nonlinear magnetosonic solitary waves. 
\par  Thus, we may conclude that the Alfv{\'e}n and magnetosonic waves get coupled and have higher-order dispersion by the Hall effect. Also, they have the same form of dispersion relation with odd (cubic or higher) order of dispersion in the long-wavelength limit, implying that both can propagate as low-frequency long-wavelength fundamental modes, whose evolution as weakly nonlinear solitary waves can be governed by KdV-like equations, however, their nonlinear couplings may not be effective by the Hall effect.
\subsection{Wave motion with the Coriolis force and Hall effects}\label{sec-sub-co-ha}
We consider a general situation in which both the Hall and Coriolis effects are present in the model. Before we study the general dispersion equation \eqref{eq-disp-eta0} in Sec. \ref{sec-results}, we consider some particular cases and obtain approximate dispersion relations for the Alfv{\'e}n and magnetosonic waves. For example, when $\alpha=\pi/2$,  Eq. \eqref{eq-disp-eta0} reduces to the the same as Eq. \eqref{eq-disp-Cor-alphapi}, i.e., the Hall contribution disappears and so is the Alfv{\'e}n mode to disappear, and  we have only the fast and slow magnetosonic modes.  However, for propagation parallel to the static magnetic field $\left(\alpha=0\right)$ , Eq. \eqref{eq-disp-eta0} reduces to
\begin{equation}\label{eq-disp-HC-alpha0}
\begin{split}
&\left(\omega^2-k^2\right)\Big\lbrace\omega^4-(1+\tilde{c}_s^2)k^2\omega^2+\tilde{c}_s^2k^4\Big\rbrace-4\Omega_0^2 \omega^2\\
&\times \left(\omega^2-\tilde{c}_s^2 k^2\sin^2{\lambda}- k^2\cos^2{\lambda}\right)-\omega^2k^4\left(\omega^2-\tilde{c}_s^2k^2\right) 
\\
& +4\Omega_0k^4\sin{\lambda}\left(\omega^2 -\tilde{c}_s^2k^2\right)+4\Omega_0^2k^4\left(\omega^2-\tilde{c}_s^2k^2\sin^2{\lambda}\right)=0.
\end{split}
\end{equation}
In the low-beta regime, $\beta\sim\tilde{c}^2_s<<1$ and if $\omega>>k\tilde{c}_s$, Eq. \eqref{eq-disp-HC-alpha0} further reduces to
\begin{equation}\label{eq-disp-HC-alpha0-1}
\begin{split}
&\left(\omega^2-k^2\right)^2-4\Omega_0^2\left(\omega^2-k^2\cos^2\lambda\right)-k^4\omega^2\\&+4\Omega_0k^4\sin\lambda+4\Omega_0^2k^4=0,
\end{split}
\end{equation}
which gives two stable kinetic Alfv{\'e}n wave modes modified by the Hall and Coriolis force effects. An approximate dispersion relation for the shear Alfv{\'e}n wave with $0<\alpha<\pi/2$ can be obtained by the same way as in Sec. \ref{sec-sub-coriolis} as 
\begin{equation}\label{eq-disp-HC-Alf}
\omega^2=d_1+d_3k^2+d_2k^4,
\end{equation}
where 
\begin{equation}\label{exp-d3}
\begin{split}
&d_3=\frac{1}{\sin^2\alpha}\left[\sin^2\alpha\cos^2\alpha+4\Omega_0\cos\alpha \right.
\\ & \left.\lbrace \cos\alpha\sin{\left(\alpha+\lambda\right)}-\tilde c_s^2\sin\lambda\rbrace+4\Omega_0^2 \right.\\ &\left.\left(\cos^2\alpha-\tilde c_s^2\sin^2\lambda\right) 
\right].
\end{split}
\end{equation}
The term proportional to $d_3$ appears due to the combined influence of the Hall and Coriolis force effects and it vanishes for $\alpha=\pi/2$ and $\lambda=0$. 
In the long-wavelength limit, $k\ll 1$, Eq. \eqref{eq-disp-HC-Alf} reduces to
\begin{equation}\label{eq-disp-HC-Alf_approx}
\omega\approx \sqrt{d_1}+\frac{d_3k^2}{2\sqrt{d_1}}+\frac{d_2k^4}{2\sqrt{d_1}},
\end{equation}
where $d_1>0$. This dispersion equation has the same form as Eq. \eqref{eq-disp-Cor-Alf} except with the higher-order dispersion $(\propto k^4)$, which appears due to the Hall effect. So, the evolution of weakly nonlinear  Alfv{\'e}n wave envelope can be described by a NLS-like equation.
By the same way, an approximate dispersion relation for the magnetosonic  wave with $0<\alpha<\pi/2$ can be obtained as  
\begin{equation}\label{eq-disp-HC-mag}
\begin{split}
&\omega^2=\frac{k}{2}\left[\left(1+\tilde c_s^2\right)k+ \right.
\\
&\left.\sqrt{b_{1\pm}+\lbrace 2a^2_\pm -\left(1+\tilde c_s^2\right)\rbrace^2k^2-b_{3\pm}k^2+b_{2\pm}k^4}\right],
\end{split}
\end{equation}
where $a_{\pm}$ is defined before in Eq. \eqref{eq-a} and $b_{3\pm}$ is given by
\begin{equation}
\begin{split}
&b_{3\pm}= \frac{16\Omega_0\cos^2\alpha}{\left(a_{\pm}^2-\cos^2{\alpha}\right)}\left[ \lbrace a_{\pm}^2\sin\left(\lambda+\alpha\right)-\tilde c_s^2 \sin \lambda \cos\alpha\rbrace \right.\\ &\left.+\Omega_0\left( a_{\pm}^2-\tilde c_s^2 \sin^2 \lambda \right)\right].
\end{split}
\end{equation} 
The term proportional to $b_{3\pm}$ appears due to the combined influence of the Hall and Coriolis force effects, and it vanishes at $\alpha=\pi/2$. In the long-wavelength limit, $k<<1$, Eq. \eqref{eq-disp-HC-mag} reduces to
\begin{equation}\label{eq-disp-HC-mag_approx}
\begin{split}
&\omega\approx\frac{b_{1\pm}^{1/4}}{\sqrt{2}}k^{1/2}+\frac{\left(1+\tilde c_s^2\right)}{2\sqrt{2}b_{1\pm}^{1/4}}k^{3/2}\\
&+\frac{\left[\lbrace 2a_{\pm}^2-\left(1+\tilde c_s^2\right)\rbrace^2-b_{3\pm}\right]}{4\sqrt{2}b_{1\pm}^{3/4}}k^{5/2}+\frac{b_{2\pm}}{4\sqrt{2}b_{1\pm}^{3/4}}k^{9/2}.
\end{split}
\end{equation}
This dispersion equation has the same form as Eq. \eqref{eq-magne-Corio} except the higher-order dispersion $(\propto k^{9/2})$, which appears due to the Hall effect. So, in the nonlinear regime, the magnetosonic mode can evolve as solitary waves, whose evolution can be described by the KdV-like equations.
\section{Results and discussion} \label{sec-results}
\begin{figure*}
\includegraphics[scale=0.5]{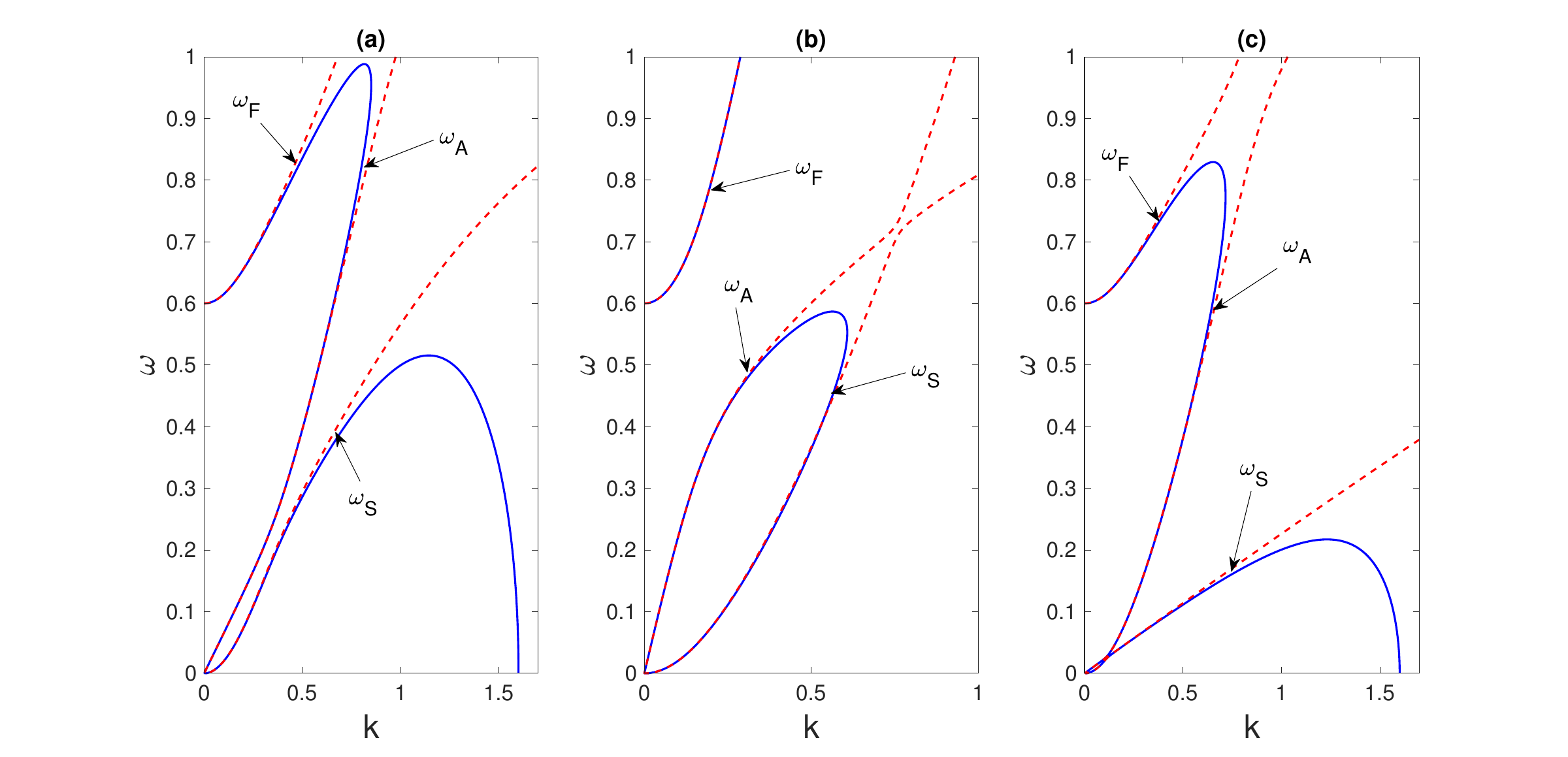}
\caption{The dispersion relation is contour plotted in the $\omega k$ plane to show three fundamental MHD  modes: the Alfv{\'e}n wave ($\omega_A$) and the Fast ($\omega_F$) and slow ($\omega_S$) magnetosonic waves, as indicated in the figure, when both the Coriolis force and the Hall effects are included in the model. The dashed and solid lines, respectively, correspond to the dispersion relations \eqref{eq-disp-eta0} and \eqref{eq-disp-eta1}. Unless otherwise stated, the fixed parameter values are $\Omega_0=0.3$, $\alpha=40^\circ$, and $\lambda=50^\circ$.  Subplots (a), (b), and (c) show the modes in three different regimes: (a) $\beta\sim1 ~(\tilde{c}_s\sim0.85)$, (b) $\beta>1~(\tilde{c}_s\sim3)$, and (c) $\beta<1~(\tilde{c}_s\sim0.3)$ respectively.}
\label{fig1}
\end{figure*}

In this section, we numerically study the dispersion characteristics of Alfv{\'e}n and magnetosonic waves and their coupling by the influences of the Coriolis force and Hall resistance, as well as the obliqueness of the propagation vector to the magnetic field and the fluid rotational angle $\lambda$ about the $y$-axis. From the general dispersion relation \eqref{eq-disp-eta0}, we observe that there are five terms, and in each term, there appear two main factors, one of which corresponds to an Alfv{\'e}n mode and the other to magnetosonic waves. Specifically, for the first term, the first factor corresponds to the  Alfv{\'e}n wave, whereas the second factor corresponds to the magnetosonic modes. We can verify it by disregarding the Coriolis force and Hall effects and equating each factor to zero. Similarly, for the second (due to the Coriolis force or rotational effect) and third (due to the Hall effect) terms, the factor $\omega^2$ may be for the Alfv{\'e}n mode and the factors in the square brackets and parentheses are for the magnetosonic modes. Inspecting the fourth and fifth terms (which appear due to combined influences of the Hall resistance and Coriolis force), we find that the factor $k^2\cos^2\alpha$ can correspond to the Alfv{\'e}n mode and the factors in the square brackets and parentheses to the magnetosonic modes. Thus, upon viewing the appearance of the Alfv{\'e}n mode factor in the fourth and fifth terms with the Hall effects, it may be reasonable to replace the factor $\omega^2$ in the third term (associated with the Hall effect) by  $k^2\cos^2\alpha$. From our numerical results, we will see that such an approximation more precisely captures the coupling between the Alfv{\'e}n and fast magnetosonic waves and the Alfv{\'e}n and slow magnetosonic waves in the forms of thumb and teardrop-like curves, respectively (See, e.g., Fig. \ref{fig1}). Here, we note that Eq. \eqref{eq-disp-eta0} also exhibits similar couplings between the modes but outside the domains: $\omega,~k<1$ and with a different set of values of $\alpha$ and $\lambda$. The advantage of considering Eq. \eqref{eq-disp-eta1} is that it exhibits coupling within the domains of definitions of $\omega$ and $k$, i.e., $\omega,~k\lesssim1$. For a similar approximation of $\omega^2$ by  $k^2\cos^2\alpha$, we may refer to Ref. \cite{ruderman2020kadomtsev}. Thus, the modeling of the dispersion equation by Eq. \eqref{eq-disp-eta1} instead of Eq. \eqref{eq-disp-eta0} may be reasonable, especially while predicting or relating to astrophysical MHD wave phenomena. So, from Eq. \eqref{eq-disp-eta0}, we obtain the following reduced dispersion relation:
\begin{widetext}
\begin{equation}\label{eq-disp-eta1}
\begin{split}
&\left(\omega^2-k^2\cos^2{\alpha}\right)\left[\omega^4-(1+\tilde{c}_s^2)k^2\omega^2+\tilde{c}_s^2k^4\cos^2{\alpha}\right]-4\Omega_0^2 \omega^2\left[\omega^2-\tilde{c}_s^2 k^2\sin^2{\lambda}- k^2\cos^2{\left(\alpha+\lambda\right)}\right] \\
& -k^6 \cos^4{\alpha} \left(\omega^2-\tilde{c}_s^2k^2\right)+4\Omega_0k^4\cos^2{\alpha}\left[\omega^2 \sin{\left(\alpha+\lambda\right)}-\tilde{c}_s^2k^2\cos{\alpha}\sin{\lambda}\right]+4\Omega_0^2k^4\cos^2{\alpha}\left(\omega^2-\tilde{c}_s^2k^2\sin^2{\lambda}\right)=0.
\end{split}
\end{equation}
\end{widetext}
\par 
We contour plot the dispersion relation \eqref{eq-disp-eta1} in three different regimes of $\beta$: (a) $\beta\sim1$, (b) $\beta>1$, and (c) $\beta<1$, to exhibit the dispersion curves for three fundamental modes, namely the Alfv{\'e}n wave ($\omega_A$) and the fast ($\omega_F$) and slow ($\omega_S$) magnetosonic modes, as well as their coupling as shown by the solid lines in Fig. \ref{fig1}. The dashed lines represent the curves corresponding to Eq. \eqref{eq-disp-eta0}.  We observe that while the coupling between the Alfv{\'e}n and fast magnetosonic waves become viable in the regimes of $\beta\lesssim1$, such as those in the solar corona and upper chromospheric regions, the same between the Alfv{\'e}n and slow magnetosonic waves occurs in high-$\beta$ ($\beta>1$) regimes, e.g., in the solar photosphere and lower chromospheric regions. Comparing the dispersion curves corresponding to Eqs. \eqref{eq-disp-eta0} and \eqref{eq-disp-eta1}, we see that the dispersion equation \eqref{eq-disp-eta1} exhibits stronger couplings in the forms of thumb- and teardrop-like curves between the modes than Eq. \eqref{eq-disp-eta0} and that the dispersion curves of Eqs. \eqref{eq-disp-eta0} and \eqref{eq-disp-eta1}  mostly agree in the propagation domains $0\lesssim(\omega,~k)\lesssim1$.  However, in contrast to Eq. \eqref{eq-disp-eta0}, Eq. \eqref{eq-disp-eta1} predicts the cut-off frequencies for the slow mode at finite values of $k$. An important point is that all the MHD waves corresponding to Eq. \eqref{eq-disp-eta1} propagate with a frequency below the ion-cyclotron frequency, and the coupling between the modes occurs within the domain $0\lesssim k\lesssim1$.   
\par 
Next, we study the influence of the Coriolis force on the wave coupling and dispersion properties of Alfv{\'e}n and magnetosonic modes using Eq. \eqref{eq-disp-eta1}. The results are shown in Fig. \ref{fig2}. In Sec. \ref{sec-sub-coriolis}, we have seen that the Coriolis force not only plays a crucial role in the wave coupling, it also retains the magnetic properties of the magnetosonic modes even when wave propagation is parallel to the magnetic field. So an analysis of how the Coriolis force influences the MHD modes is pertinent. From subplot (a), we see that as the angular velocity $\Omega_0$  is reduced, or the contribution from the Coriolis force weakens, the thumb curve splits, resulting in separation of the fast magnetosonic and Alfv{\'e}n modes within the frequency domain $(0<\omega<1)$ and exhibiting weak coupling between the modes, i.e., larger the Coriolis force, the stronger is the coupling between the fast magnetosonic and the Alfv{\'e}n modes, especially in the regime $\beta\sim1$. However, the values of $\Omega_0$ should be limited such that the values of $\omega$ do not greatly exceed unity.  In the regime of $\beta\lesssim1$, subplots (a) and (c) show that while the fast magnetosonic mode frequency is significantly reduced, the frequency of the Alfv{\'e}n wave remains almost unchanged with a reduction of $\Omega_0$. In the latter at $\Omega_0=0.2$, the frequency of the slow magnetosonic mode also gets reduced but at higher values of $k$ having a cut-off at a lower value of $k$ than when $\Omega_0=0.3$ and its coupling with the Alfv{\'e}n mode remains stronger in the form of a thumb curve.  On the other hand, in the high-$\beta$ regime with $\beta>1$, the fast magnetosonic mode has a similar property with a wave-frequency reduction as in the cases $\beta\lesssim1$ [subplots (a) and (c)]. However, while the  Alfv{\'e}n wave frequency gets reduced, the frequency of the slow mode gets enhanced. Both the modes have stronger coupling, exhibiting a teardrop-like curve even with a reduction of $\Omega_0$ from $\Omega_0=0.3$ to $\Omega_0=0.2$.   From the analysis, we may conclude that to have strong coupling between the Alfv{\'e}n and fast magnetosonic modes in the regime of $\beta\sim1$, the values of $\Omega_0$ must exceed a critical value. For the coupling of the other modes in other regimes of $\beta$, there is no such critical value. 
\begin{figure*}
\includegraphics[scale=0.5]{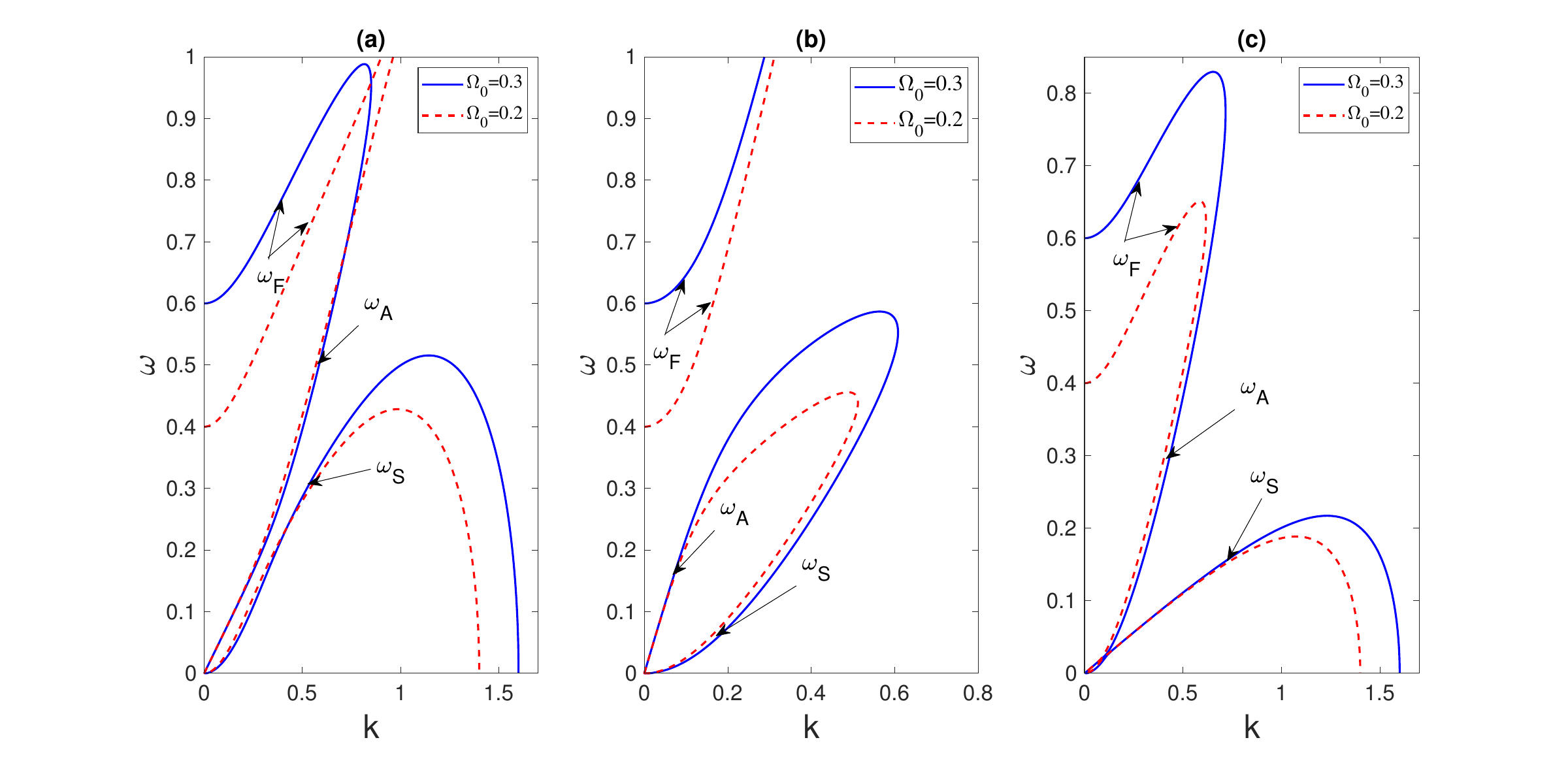}
\caption{The dispersion relation [Eq. \eqref{eq-disp-eta1}] is contour plotted in the $\omega k$ plane to show three fundamental MHD  modes: Alfv{\'e}n wave ($\omega_A$) and Fast ($\omega_F$) and slow ($\omega_S$) magnetosonic waves as indicated in the figure when both the Coriolis force and the Hall effects are included in the model. The solid and dashed lines, respectively, correspond to two different values of $\Omega_0$: $\Omega_0=0.3$ and $\Omega_0=0.2$, or as in the legends. Unless otherwise stated, the fixed parameter values are $\alpha=40^\circ$ and $\lambda=50^\circ$.  Subplots (a), (b), and (c) show the modes in three different regimes: (a) $\beta\sim1 ~(\tilde{c}_s=0.85)$, (b) $\beta>1~(\tilde{c}_s=3)$, and (c) $\beta<1~(\tilde{c}_s=0.3)$, respectively.}
\end{figure*}
\par
\begin{figure*}
\includegraphics[scale=0.5]{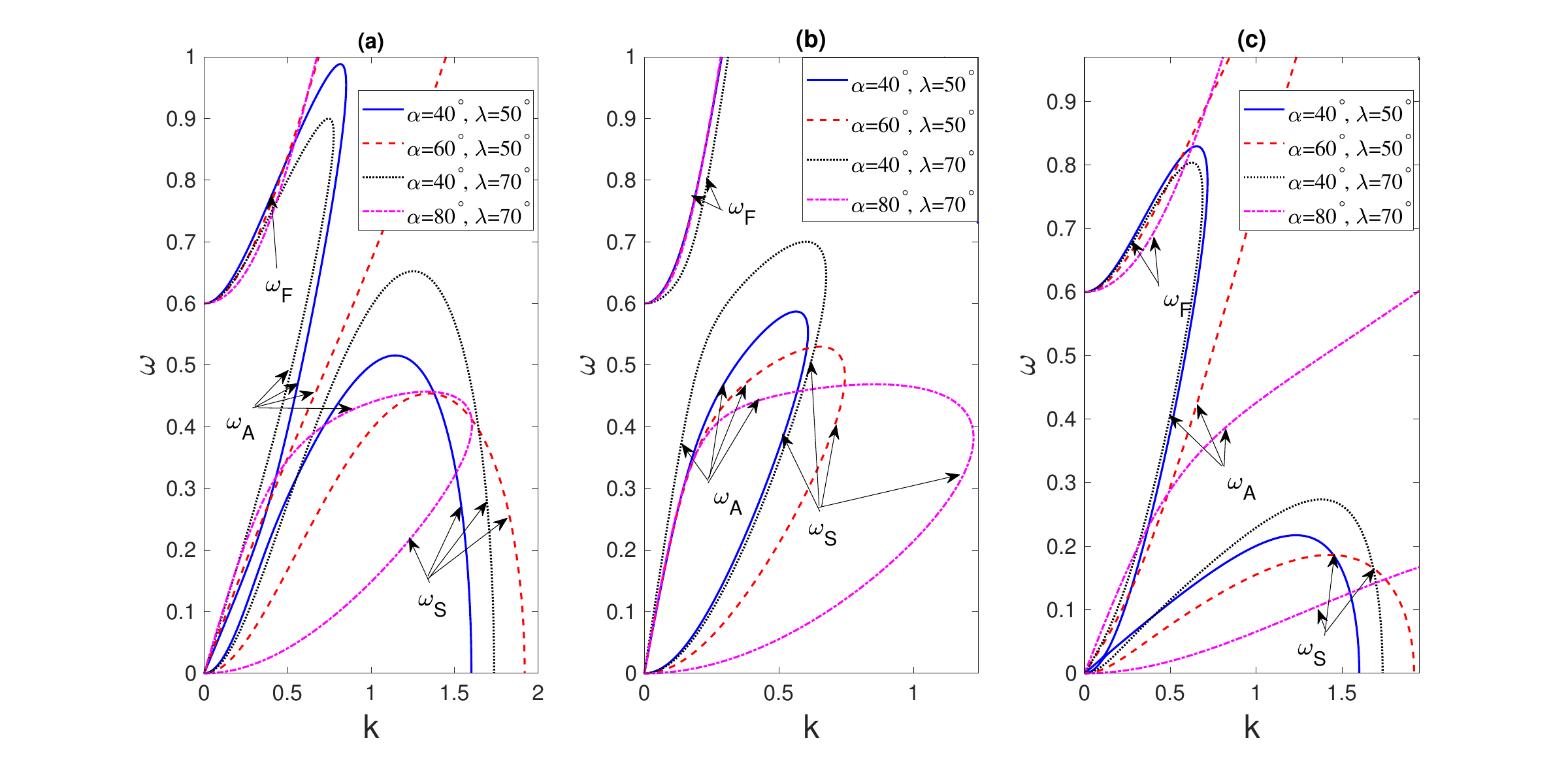}
\caption{The dispersion relation [Eq. \eqref{eq-disp-eta1}] is contour plotted in the $\omega k$ plane to show three fundamental MHD  modes: Alfv{\'e}n wave ($\omega_A$) and Fast ($\omega_F$) and slow ($\omega_S$) magnetosonic waves as indicated in the figure when both the Coriolis force and the Hall effects are included in the model. The solid, dashed, and dotted lines, respectively, correspond to different values of the angles $\alpha$ (made by the external magnetic field with the $x$-axis) and $\lambda$ (made by the axis of rotation with the $y$-axis) as indicated in the legends but with a fixed $\Omega_0=0.3$.  Subplots (a), (b), and (c) show the modes in three different regimes: (a) $\beta\sim1 ~(\tilde{c}_s=0.85)$, (b) $\beta>1~(\tilde{c}_s=3)$, and (c) $\beta<1~(\tilde{c}_s=0.3)$, respectively. }
\label{fig3}
\end{figure*}

 
We also study the effects of the angle of propagation $\alpha$ and the rotational angle $\lambda$ on the dispersion properties of the MHD waves and their coupling.  The results are displayed in Fig. \ref{fig3}. From subplots (a) and (c) (in the regimes of $\beta\sim1$ and $\beta<1$), it is interesting to note that depending on the values of $\lambda$ and $\alpha$, the thumb curve for the coupling of the fast magnetosonic mode and the Alfv{\'e}n wave may form or split into two separate curves. For example, for subplot (a), the thumb curve forms (exhibiting strong coupling) in the regimes: $0<\alpha\lesssim40^\circ$, $0<\lambda<90^\circ$; $\alpha\approx45^\circ$, $0<\lambda<65^\circ$, and $\alpha\approx50^\circ$, $0<\lambda<25^\circ$. For the subplot (c), the regimes are $0<\alpha\lesssim59^\circ$, $0<\lambda<25^\circ$; $\alpha\approx60^\circ$, $75^\circ<\lambda<90^\circ$; and $\alpha\approx61^\circ$, $80<\lambda<90^\circ$, otherwise the thumb curve splits, resulting a weak coupling between the fast magnetosonic and the Alfv{\'e}n modes. On the other hand, subplot (b) shows that in the regimes of $\beta>1$, the coupling between the Alfv{\'e}n and slow magnetosonic modes occur in the form of a teardrop curve, which remains for wide ranges of values of $\alpha$ and $\lambda$ in $0<(\alpha,~\lambda)<90^\circ$.   
 A novel feature in the regime of $\beta\sim1$ [subplot (a)] is that a switching of the coupling occurs, i.e., the thumb curve coupling between the fast magnetosonic and Alfv{\'e}n modes [which is also the case in the low-$\beta$ regime with $\beta<1$; See subplot (c)] shifts to the teardrop coupling between the Alfv{\'e}n and slow magnetosonic modes [which is typically the case in the high-$\beta$ regime with $\beta>1$; See subplot (b)] when both $\alpha$ and $\lambda$ assume values close to $70^\circ$ or more, e.g., in the regimes  $68^\circ\lesssim\alpha<90^\circ$, $72^\circ\lesssim\lambda<90^\circ$ and $69^\circ\lesssim\alpha<90^\circ$, $68^\circ\lesssim\lambda<90^\circ$.  From subplots (a) and (c), we also observe that unless $\alpha$ and $\lambda$ assume values close to $70^\circ$ or more, the slow magnetosonic mode exhibits cut-offs at different finite values of the wave number $k$ as the parameters $\alpha$ and $\lambda$ vary.  
\par
\begin{figure*}
\includegraphics[scale=0.5]{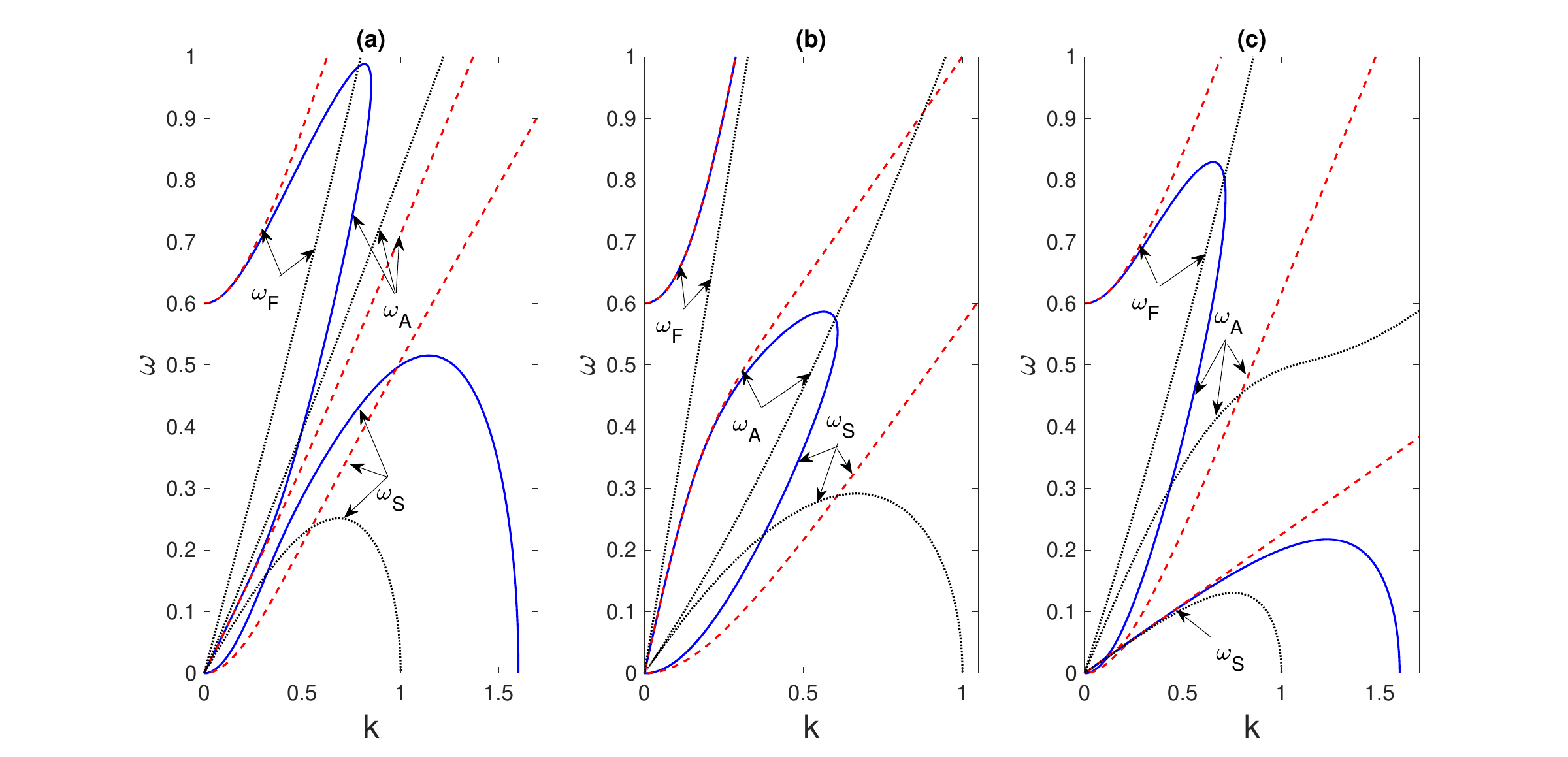}
\caption{The dispersion relation [Eq. \eqref{eq-disp-eta1}] is contour plotted in the $\omega k$ plane to show three fundamental MHD  modes: Alfv{\'e}n wave ($\omega_A$) and Fast ($\omega_F$) and slow ($\omega_S$) magnetosonic waves as indicated in the figure. The solid, dashed and dotted lines, respectively, correspond to the cases when (i) both the Coriolis force (with $\Omega_0=0.3$) and Hall  effects are present, (ii) the Coriolis force is present (with $\Omega_0=0.3$) but in absence of the Hall effects, (iii) the Hall effects are included but in absence of the Coriolis force. The other fixed parameter values are $\alpha=40^\circ$ and $\lambda=50^\circ$. Subplots (a), (b), and (c) show the modes in three different regimes: (a) $\beta\sim1 ~(\tilde{c}_s=0.85)$, (b) $\beta>1~(\tilde{c}_s=3)$, and (c) $\beta<1~(\tilde{c}_s=0.3)$ respectively.   }
\label{fig4}
\end{figure*}

 
Finally, we study the relative or combined influence of the Coriolis force and Hall effects on the dispersion characteristics of MHD waves. The results are shown in Fig. \ref{fig4} in three different regimes of $\beta$: (a) $\beta\sim1$, (b) $\beta>1$, and (c) $\beta<1$. We consider the cases: (i) when both the effects are present and (ii) when any of them is present. Comparing the solid lines with the dashed or dotted lines, we observe that the strong coupling between the fast magnetosonic and Alfv{\'e}n modes (thumb curve) and  between the Alfv{\'e}n and slow magnetosonic modes (teardrop curve) are possible only when both the effects are present in the model. Also, from the dashed and dotted lines, we note that the slow magnetosonic modes exhibit cut-offs at finite values of $k$ by the influence of the Hall effect.  Thus, we remark that the combined influence of the Coriolis force and the Hall effect is the prerequisite for the strong coupling of Alfv{\'e}n and magnetosonic modes in which the slow mode exhibits cut-off by the influence of the Hall resistance and the rotational frequency should be above a critical value but be limited $(<1)$ such that the centrifugal force may remain smaller than the Coriolis force.     
\section{Applications to Solar and fusion plasmas} \label{sec-appl}
The Sun's atmosphere, which typically consists of the photosphere, chromosphere, and corona, is an active medium with a wide range of temperatures and densities. Due to, e.g., convective gas motions, the three fundamental MHD modes, namely, the Alfv{\'e}n wave and the fast and slow magnetosonic modes, can be excited in the medium. Typically, in a uniform plasma, these modes propagate independently. However, in an inhomogeneous medium (e.g., coronal plasma), they propagate as coupled modes. We have seen in Secs. \ref{sec-sub-coriolis}-\ref{sec-sub-co-ha} that even in a homogeneous medium, the coupling between the modes can be possible by the influence of either the Coriolis force or the Hall resistance effect, and the coupling can be strong by their combined influences. While the solar photosphere and chromosphere are typically partially ionized plasmas, the highly ionized corona can be considered as collisionless and its dynamics can be described by an ideal MHD or a HMHD model \cite{pandey2008hall}. Furthermore, ground- and space-based observations have confirmed small- ($3-10$ km/s) as well as large-amplitude (more than $10$ km/s) oscillatory motions in the Solar atmosphere, which have been interpreted in terms of propagating MHD modes. 
\par 
In what follows, we study the relevance of the MHD modes, especially the Hall and Coriolis force effects in the Solar corona. To this end, we consider typical plasma parameters and examine the Hall time and length scales, together with the rotational angle and the angular velocity associated with the Coriolis force. Coronal heating by MHD waves has been a feasible mechanism for interpreting high coronal temperatures compared to the lower photosphere. In the latter, the MHD waves are excited by the footpoint motion of the magnetic field lines, which later emerge into the corona. In this context, Alfv{\'e}n waves, while propagating in the corona, may lose energy to plasma particles by resonance damping and thus can accelerate solar plasmas from coronal holes. The power spectra of horizontal motions in the photosphere also indicate the existence of MHD waves with frequencies ranging from $10^{-5}$ to $0.1$ Hz at a few solar radii. So, one can expect an observable frequency of the order of $1$ Hz. Also, low-frequency whistler waves can exist close to the footprint of the magnetic flux with reduced neutral density. These waves later turn into Alfv{\'e}n waves, which propagate into the corona and heat it. Due to the decreasing neutral density and increasing ionization, such waves propagate almost undamped in the corona. Thus, the Hall effect may be responsible for exciting whistler waves in the lower photosphere or lower part of the corona. Typically, the Hall scale size is approximately more than a few tens of kms. Thus, the MHD modes due to the Hall effects could be important for modeling various coronal heating. 
\par 
In the present fully ionized plasma model, the Hall time scale is $\tau_H\sim \omega_{ci}^{-1}$, i.e., the inverse of the ion-cyclotron frequency and the Hall length-scale is $L_H\sim V_A/\omega_{ci}\equiv\lambda_i$, i.e., the ion skin depth, implying that in fully ionized plasmas (e.g., Solar corona), the Hall effects become important for $\omega\gtrsim\omega_H\sim\omega_{ci}$, or $\omega\lesssim\omega_{ci}$ and the length-scale comparable to the ion skin depth, i.e., $L_H\sim\lambda_i$. For typical parameters (relevant for coronal loops) with the magnetic field strength $B_0\sim100$ G, the temperature $T\sim6\times10^6$ K, the number density $n_0\sim1.4\times10^9~\rm{cm}^{-3}$, and assuming the ion mass as the proton mass, we have the Alfv{\'e}n speed $V_A\sim6\times10^6$ m/s, the ion skin depth or the Hall length-scale $L_H\sim6$ m, and the Hall frequency or the ion-cyclotron frequency $\omega_H\sim\omega_{ci}\sim9.7\times10^5~\rm{s}^{-1}$. Adopting a coronal region of radius $a\sim10$ m, or $1$ m, one can have the Alfv{\'e}n frequency as $\omega_A=a/V_A\sim6\times10^5~\rm{s}^{-1}$, or $6\times10^6~\rm{s}^{-1}$, i.e., the Hall effects are important either for $\omega\lesssim\omega_{ci}$, or for $\omega\gtrsim\omega_{ci}$. While in the former, the Hall effect is comparable to the Coriolis force effect since $\omega_A\sim2\Omega_0\sim6\times10^5~\rm{s}^{-1}$, in the latter, the Hall effects can dominate over the Coriolis force. It also follows that the Hall frequency is much higher but the Hall length-scale is  much lower than the corresponding scales for the solar photosphere at an altitude of about $10^3$ km or ionospheric plasmas at an altitude of about $150$ km (or lower). 
\par
On the other hand, Hall effects can play crucial roles in tokamak discharges or non-Ohmic current drive schemes \cite{pandey1995hall}. They are also important near the wall region of a tokamak . Since near the wall region, the ionization rate is high $(n_n/n_i\sim10^{-4}-10^{-3})$ with $n_n$ denoting the neutral density, the Hall time and length scales will remain the same as for fully ionized coronal plasmas.  For typical fusion plasma parameters \cite{pandey2008hall} with         
$B_0\sim10$ kG, $n_0\sim1.4\times10^9~\rm{cm}^{-3}$, and assuming the ion mass as the proton mass, we have $V_A\sim2.2\times10^6$ m/s, $L_H\sim\lambda_i\sim0.023$ m, and $\omega_H\sim\omega_{ci}\sim9.7\times10^7~\rm{s}^{-1}$. Adopting a major radius of the tokamak, $a\sim 1$ m, we have $\omega_A\sim2.2\times10^{6}~\rm{s}^{-1}$, i.e., the Hall scales are much smaller than coronal plasmas. Thus, in fusion plasmas (near the wall region of a tokamak), the Hall effects may be important for $\omega\sim\omega_{ci}$ and the length-scale comparable to the ion skin depth, i.e., $L_H\sim\lambda_i$.
\section{Summary and conclusion}\label{Sec-Conclusion}
We have studied the existence and coupling of obliquely propagating Alfv{\'e}n and magnetosonic waves in a rotating Hall-MHD magnetoplasma, as well as analyzed the characteristics of these modes in three different regimes of $\beta$ by the influences of the obliqueness of wave propagation, the angular velocity and rotational angle associated with the Coriolis force, and the Hall resistance effect. We have seen that the coupling between the Alfv{\'e}n and magnetosonic waves can occur by the Coriolis force or Hall effects. Also, while in the regime $\beta\lesssim1$, the coupling between the fast magnetosonic and Alfv{\'e}n modes become prominent and they appear in the form of a thumb curve, the teardrop-like coupling between the slow  magnetosonic and Alfv{\'e}n modes occurs in the high-$\beta$ $(\beta>1)$ regime.    
We summarize the main results as follows:
\begin{itemize}
\item The MHD waves propagate with a frequency below the ion-cyclotron frequency, and the coupling between the modes occurs within the domain $0\lesssim k\lesssim1$.
\item{\textit{Without Hall effect:}} In the absence of the Hall effect, an interesting feature in a rotating plasma is that even for MHD wave propagation parallel to the static magnetic field, not only are both the magnetic and acoustic characters of the magnetosonic modes retained, the modes also get coupled to the Alfv{\'e}n mode by the influence of the Coriolis force. However, in the absence of the Coriolis force, both the Alfv{\'e}n and magnetosonic modes get decoupled, giving the fast mode to appear as the Alfv{\'e}n mode: $\omega=k$ and the slow mode as the acoustic mode: $\omega=\tilde{c}_s k$. 
\item \textit{Without Hall effect:} The Alfv{\'e}n and magnetosonic waves become dispersive due to the Coriolis force effect with even $(\geq2)$ and odd $(\geq3)$ orders of dispersion, implying that their  nonlinear evolution as solitary waves can be described by NLS- and KdV-like equations, respectively, and the nonlinear coupling of these waves (to be governed by a coupled KdV- and NLS-like equations) could be more pronounced in the presence of the Coriolis force.
\item \textit{Without Coriolis force:} In the absence of the Coriolis force (non-rotating plasmas), the Alfv{\'e}n and magnetosonic waves also get coupled and have higher-order dispersion (than by the Coriolis force) by the Hall effect. Also, they have the same form of dispersion relation with odd (cubic or higher) order of dispersion in the long-wavelength limit, implying that both can propagate as low-frequency long-wavelength fundamental modes, whose evolution as weakly nonlinear solitary waves can be governed by KdV-like equations, however, their nonlinear couplings may not be effective by the Hall effect.
\item \textit{With Coriolis force and Hall effect:} To have strong coupling between the Alfv{\'e}n and fast magnetosonic modes in the regime of $\beta\sim1$, the values of $\Omega_0$ must exceed a critical value. For the coupling of the other modes in other regimes of $\beta$, there is no such critical value. In the regime, $\beta\sim1$, a switching of the coupling occurs, i.e., the thumb curve coupling between the fast magnetosonic and Alfv{\'e}n modes (which is the case for $\beta<1$) shifts to the teardrop coupling between the Alfv{\'e}n and slow magnetosonic modes (which is typically the case for $\beta>1$) when both $\alpha$ and $\lambda$ assume values close to $70^\circ$ or more (but less than $90^\circ$).
 \end{itemize}
 To conclude, the combined influence of the Coriolis force and the Hall effect is the prerequisite for the strong coupling of Alfv{\'e}n and magnetosonic modes in which the slow mode exhibits cut-off by the influence of the Hall resistance and the rotational frequency should be above a critical value but be limited $(<\omega_{ci})$ such that the centrifugal force remains smaller than the Coriolis force. We have noted that the coupling between the slow magnetosonic and Alfv{\'e}n modes is typically weak (except in a particular domain of $\Omega_0$ stated above; \textit{cf}. Fig. \ref{fig4}) and that between the fast magnetosonic and Alfv{\'e}n modes is rather strong in the low-$\beta$ regime. Since Alfv{\'e}n wave can drive both the slow and fast modes, the process of driving the fast modes can be enhanced by the phase mixing effect due to magnetic field inhomogeneity. Furthermore, it has been shown that the inhomogeneities in the pressure gradient and magnetic field-line curvature can play crucial roles in the MHD wave couplings \cite{ohtani1989}. So, considering these inhomogeneity effects in the model could enrich the coupling mechanism and modify their dispersion characteristics, which could be a project for future research. Also, in the present HMHD model, we have neglected the electron inertial effect. However, since electrons can drift more freely than ions with the magnetic field, such effects could be important at frequencies higher than the ion-cyclotron frequency. So, extending the HMHD with electron inertia included could be another project of future study. 
\section*{Acknowledgments} The authors gratefully acknowledge helpful comments from Birendra Pandey of Macquarie University, Australia, during the preparation of this work. 
\section*{Author declarations}
\subsection*{Conflict of interest}  The authors have no conflicts to disclose.
\subsection*{Author Contributions} \textbf{Amar P.  Misra:} Conceptualization (lead); Investigation (lead); Formal analysis (equal);  Methodology (equal); Supervision (lead); Validation (equal); Writing--original draft (lead). 
 \textbf{Rupak Dey:} Conceptualization (supporting); Investigation (equal); Formal analysis (equal);  Methodology (equal); Validation (equal); Writing--original draft (equal). 
\textbf{Vinod Krishan:} Conceptualization (supporting); Investigation (supporting); Formal analysis (supporting); Methodology (supporting); Validation (equal); Writing--review \& editing (supporting).

 \section*{Data availability}
 The data that support the findings of this study are available from  the corresponding author upon reasonable request.
\bibliography{References}

\begin{thebibliography}{32}%
\makeatletter
\providecommand \@ifxundefined [1]{%
 \@ifx{#1\undefined}
}%
\providecommand \@ifnum [1]{%
 \ifnum #1\expandafter \@firstoftwo
 \else \expandafter \@secondoftwo
 \fi
}%
\providecommand \@ifx [1]{%
 \ifx #1\expandafter \@firstoftwo
 \else \expandafter \@secondoftwo
 \fi
}%
\providecommand \natexlab [1]{#1}%
\providecommand \enquote  [1]{``#1''}%
\providecommand \bibnamefont  [1]{#1}%
\providecommand \bibfnamefont [1]{#1}%
\providecommand \citenamefont [1]{#1}%
\providecommand \href@noop [0]{\@secondoftwo}%
\providecommand \href [0]{\begingroup \@sanitize@url \@href}%
\providecommand \@href[1]{\@@startlink{#1}\@@href}%
\providecommand \@@href[1]{\endgroup#1\@@endlink}%
\providecommand \@sanitize@url [0]{\catcode `\\12\catcode `\$12\catcode
  `\&12\catcode `\#12\catcode `\^12\catcode `\_12\catcode `\%12\relax}%
\providecommand \@@startlink[1]{}%
\providecommand \@@endlink[0]{}%
\providecommand \url  [0]{\begingroup\@sanitize@url \@url }%
\providecommand \@url [1]{\endgroup\@href {#1}{\urlprefix }}%
\providecommand \urlprefix  [0]{URL }%
\providecommand \Eprint [0]{\href }%
\providecommand \doibase [0]{https://doi.org/}%
\providecommand \selectlanguage [0]{\@gobble}%
\providecommand \bibinfo  [0]{\@secondoftwo}%
\providecommand \bibfield  [0]{\@secondoftwo}%
\providecommand \translation [1]{[#1]}%
\providecommand \BibitemOpen [0]{}%
\providecommand \bibitemStop [0]{}%
\providecommand \bibitemNoStop [0]{.\EOS\space}%
\providecommand \EOS [0]{\spacefactor3000\relax}%
\providecommand \BibitemShut  [1]{\csname bibitem#1\endcsname}%
\let\auto@bib@innerbib\@empty
\bibitem [{\citenamefont {Skiathas}\ and\ \citenamefont
  {Gourgouliatos}(2024)}]{skiathas2024combined}%
  \BibitemOpen
  \bibfield  {author} {\bibinfo {author} {\bibfnamefont {D.}~\bibnamefont
  {Skiathas}}\ and\ \bibinfo {author} {\bibfnamefont {K.~N.}\ \bibnamefont
  {Gourgouliatos}},\ }\bibfield  {title} {\bibinfo {title} {{Combined magnetic
  field evolution in neutron star cores and crusts: ambipolar diffusion, Hall
  effect, and Ohmic dissipation}},\ }\href
  {https://doi.org/10.1093/mnras/stae190} {\bibfield  {journal} {\bibinfo
  {journal} {Monthly Notices of the Royal Astronomical Society}\ }\textbf
  {\bibinfo {volume} {528}},\ \bibinfo {pages} {5178} (\bibinfo {year}
  {2024})}\BibitemShut {NoStop}%
\bibitem [{\citenamefont {Pandey}\ and\ \citenamefont
  {Wardle}(2013)}]{pandey2013magnetic}%
  \BibitemOpen
  \bibfield  {author} {\bibinfo {author} {\bibfnamefont {B.~P.}\ \bibnamefont
  {Pandey}}\ and\ \bibinfo {author} {\bibfnamefont {M.}~\bibnamefont
  {Wardle}},\ }\bibfield  {title} {\bibinfo {title} {{Magnetic-diffusion-driven
  shear instability of solar flux tubes}},\ }\href
  {https://doi.org/10.1093/mnras/stt184} {\bibfield  {journal} {\bibinfo
  {journal} {Monthly Notices of the Royal Astronomical Society}\ }\textbf
  {\bibinfo {volume} {431}},\ \bibinfo {pages} {570} (\bibinfo {year}
  {2013})}\BibitemShut {NoStop}%
\bibitem [{\citenamefont {Miteva}\ \emph {et~al.}(2003)\citenamefont {Miteva},
  \citenamefont {Zhelyazkov},\ and\ \citenamefont
  {Erdélyi}}]{miteva2003surface}%
  \BibitemOpen
  \bibfield  {author} {\bibinfo {author} {\bibfnamefont {R.}~\bibnamefont
  {Miteva}}, \bibinfo {author} {\bibfnamefont {I.}~\bibnamefont {Zhelyazkov}},\
  and\ \bibinfo {author} {\bibfnamefont {R.}~\bibnamefont {Erdélyi}},\
  }\bibfield  {title} {\bibinfo {title} {{Surface wave propagation in steady
  ideal Hall-magnetohydrodynamic magnetic slabs}},\ }\href
  {https://doi.org/10.1063/1.1615769} {\bibfield  {journal} {\bibinfo
  {journal} {Physics of Plasmas}\ }\textbf {\bibinfo {volume} {10}},\ \bibinfo
  {pages} {4463} (\bibinfo {year} {2003})}\BibitemShut {NoStop}%
\bibitem [{\citenamefont {Aburjania}\ \emph {et~al.}(2005)\citenamefont
  {Aburjania}, \citenamefont {Chargazia}, \citenamefont {Jandieri},
  \citenamefont {Khantadze}, \citenamefont {Kharshiladze},\ and\ \citenamefont
  {Lominadze}}]{aburjania2005generation}%
  \BibitemOpen
  \bibfield  {author} {\bibinfo {author} {\bibfnamefont {G.}~\bibnamefont
  {Aburjania}}, \bibinfo {author} {\bibfnamefont {K.}~\bibnamefont
  {Chargazia}}, \bibinfo {author} {\bibfnamefont {G.}~\bibnamefont {Jandieri}},
  \bibinfo {author} {\bibfnamefont {A.}~\bibnamefont {Khantadze}}, \bibinfo
  {author} {\bibfnamefont {O.}~\bibnamefont {Kharshiladze}},\ and\ \bibinfo
  {author} {\bibfnamefont {J.}~\bibnamefont {Lominadze}},\ }\bibfield  {title}
  {\bibinfo {title} {Generation and propagation of the ulf planetary-scale
  electromagnetic wavy structures in the ionosphere},\ }\href
  {https://doi.org/https://doi.org/10.1016/j.pss.2005.02.004} {\bibfield
  {journal} {\bibinfo  {journal} {Planetary and Space Science}\ }\textbf
  {\bibinfo {volume} {53}},\ \bibinfo {pages} {881} (\bibinfo {year}
  {2005})}\BibitemShut {NoStop}%
\bibitem [{\citenamefont {Pandey}\ \emph {et~al.}(1995)\citenamefont {Pandey},
  \citenamefont {Avinash}, \citenamefont {Kaw},\ and\ \citenamefont
  {Sen}}]{pandey1995hall}%
  \BibitemOpen
  \bibfield  {author} {\bibinfo {author} {\bibfnamefont {B.~P.}\ \bibnamefont
  {Pandey}}, \bibinfo {author} {\bibfnamefont {K.}~\bibnamefont {Avinash}},
  \bibinfo {author} {\bibfnamefont {P.~K.}\ \bibnamefont {Kaw}},\ and\ \bibinfo
  {author} {\bibfnamefont {A.}~\bibnamefont {Sen}},\ }\bibfield  {title}
  {\bibinfo {title} {{Hall term enhancement of current drive in a bounded
  inhomogeneous plasma}},\ }\href {https://doi.org/10.1063/1.871414} {\bibfield
   {journal} {\bibinfo  {journal} {Physics of Plasmas}\ }\textbf {\bibinfo
  {volume} {2}},\ \bibinfo {pages} {629} (\bibinfo {year} {1995})}\BibitemShut
  {NoStop}%
\bibitem [{\citenamefont {Lingam}\ and\ \citenamefont
  {Bhattacharjee}(2016)}]{lingam2016hall}%
  \BibitemOpen
  \bibfield  {author} {\bibinfo {author} {\bibfnamefont {M.}~\bibnamefont
  {Lingam}}\ and\ \bibinfo {author} {\bibfnamefont {A.}~\bibnamefont
  {Bhattacharjee}},\ }\bibfield  {title} {\bibinfo {title} {Hall current
  effects in mean-field dynamo theory},\ }\href
  {https://doi.org/10.3847/0004-637X/829/1/51} {\bibfield  {journal} {\bibinfo
  {journal} {The Astrophysical Journal}\ }\textbf {\bibinfo {volume} {829}},\
  \bibinfo {pages} {51} (\bibinfo {year} {2016})}\BibitemShut {NoStop}%
\bibitem [{\citenamefont {Morales}\ \emph {et~al.}(2005)\citenamefont
  {Morales}, \citenamefont {Dasso},\ and\ \citenamefont
  {Gómez}}]{morales2005hall}%
  \BibitemOpen
  \bibfield  {author} {\bibinfo {author} {\bibfnamefont {L.~F.}\ \bibnamefont
  {Morales}}, \bibinfo {author} {\bibfnamefont {S.}~\bibnamefont {Dasso}},\
  and\ \bibinfo {author} {\bibfnamefont {D.~O.}\ \bibnamefont {Gómez}},\
  }\bibfield  {title} {\bibinfo {title} {Hall effect in incompressible magnetic
  reconnection},\ }\href {https://doi.org/https://doi.org/10.1029/2004JA010675}
  {\bibfield  {journal} {\bibinfo  {journal} {Journal of Geophysical Research:
  Space Physics}\ }\textbf {\bibinfo {volume} {110}} (\bibinfo {year}
  {2005})}\BibitemShut {NoStop}%
\bibitem [{\citenamefont {Bai}\ and\ \citenamefont
  {Stone}(2017)}]{bai2017hall}%
  \BibitemOpen
  \bibfield  {author} {\bibinfo {author} {\bibfnamefont {X.-N.}\ \bibnamefont
  {Bai}}\ and\ \bibinfo {author} {\bibfnamefont {J.~M.}\ \bibnamefont
  {Stone}},\ }\bibfield  {title} {\bibinfo {title} {Hall effect–mediated
  magnetic flux transport in protoplanetary disks},\ }\href
  {https://doi.org/10.3847/1538-4357/836/1/46} {\bibfield  {journal} {\bibinfo
  {journal} {The Astrophysical Journal}\ }\textbf {\bibinfo {volume} {836}},\
  \bibinfo {pages} {46} (\bibinfo {year} {2017})}\BibitemShut {NoStop}%
\bibitem [{\citenamefont {Alfv{\'e}n}(1942)}]{alfven1942existence}%
  \BibitemOpen
  \bibfield  {author} {\bibinfo {author} {\bibfnamefont {H.}~\bibnamefont
  {Alfv{\'e}n}},\ }\bibfield  {title} {\bibinfo {title} {Existence of
  electromagnetic-hydrodynamic waves},\ }\href
  {https://doi.org/10.1038/150405d0} {\bibfield  {journal} {\bibinfo  {journal}
  {Nature}\ }\textbf {\bibinfo {volume} {150}},\ \bibinfo {pages} {405}
  (\bibinfo {year} {1942})}\BibitemShut {NoStop}%
\bibitem [{\citenamefont {Lighthill}(1960)}]{lighthill1960studies}%
  \BibitemOpen
  \bibfield  {author} {\bibinfo {author} {\bibfnamefont {M.~J.}\ \bibnamefont
  {Lighthill}},\ }\bibfield  {title} {\bibinfo {title} {Studies on
  magneto-hydrodynamic waves and other anisotropic wave motions},\ }\href
  {https://doi.org/10.1098/rsta.1960.0010} {\bibfield  {journal} {\bibinfo
  {journal} {Philosophical Transactions of the Royal Society of London. Series
  A, Mathematical and Physical Sciences}\ }\textbf {\bibinfo {volume} {252}},\
  \bibinfo {pages} {397} (\bibinfo {year} {1960})}\BibitemShut {NoStop}%
\bibitem [{\citenamefont {Ruderman}(2020)}]{ruderman2020kadomtsev}%
  \BibitemOpen
  \bibfield  {author} {\bibinfo {author} {\bibfnamefont {M.~S.}\ \bibnamefont
  {Ruderman}},\ }\bibfield  {title} {\bibinfo {title} {Kadomtsev-petviashvili
  equation for magnetosonic waves in hall plasmas and soliton stability},\
  }\href {https://doi.org/10.1088/1402-4896/aba3a9} {\bibfield  {journal}
  {\bibinfo  {journal} {Physica Scripta}\ }\textbf {\bibinfo {volume} {95}},\
  \bibinfo {pages} {095601} (\bibinfo {year} {2020})}\BibitemShut {NoStop}%
\bibitem [{\citenamefont {Almaguer}(1992)}]{almaguer1992linear}%
  \BibitemOpen
  \bibfield  {author} {\bibinfo {author} {\bibfnamefont {J.~A.}\ \bibnamefont
  {Almaguer}},\ }\bibfield  {title} {\bibinfo {title} {{Linear waves in a
  resistive plasma with Hall current}},\ }\href
  {https://doi.org/10.1063/1.860401} {\bibfield  {journal} {\bibinfo  {journal}
  {Physics of Fluids B: Plasma Physics}\ }\textbf {\bibinfo {volume} {4}},\
  \bibinfo {pages} {3443} (\bibinfo {year} {1992})}\BibitemShut {NoStop}%
\bibitem [{\citenamefont {Pandey}\ and\ \citenamefont
  {Wardle}(2008)}]{pandey2008hall}%
  \BibitemOpen
  \bibfield  {author} {\bibinfo {author} {\bibfnamefont {B.~P.}\ \bibnamefont
  {Pandey}}\ and\ \bibinfo {author} {\bibfnamefont {M.}~\bibnamefont
  {Wardle}},\ }\bibfield  {title} {\bibinfo {title} {{Hall magnetohydrodynamics
  of partially ionized plasmas}},\ }\href
  {https://doi.org/10.1111/j.1365-2966.2008.12998.x} {\bibfield  {journal}
  {\bibinfo  {journal} {Monthly Notices of the Royal Astronomical Society}\
  }\textbf {\bibinfo {volume} {385}},\ \bibinfo {pages} {2269} (\bibinfo {year}
  {2008})}\BibitemShut {NoStop}%
\bibitem [{\citenamefont {Kawazura}(2017)}]{kawazura2017modification}%
  \BibitemOpen
  \bibfield  {author} {\bibinfo {author} {\bibfnamefont {Y.}~\bibnamefont
  {Kawazura}},\ }\bibfield  {title} {\bibinfo {title} {Modification of
  magnetohydrodynamic waves by the relativistic hall effect},\ }\href
  {https://doi.org/10.1103/PhysRevE.96.013207} {\bibfield  {journal} {\bibinfo
  {journal} {Phys. Rev. E}\ }\textbf {\bibinfo {volume} {96}},\ \bibinfo
  {pages} {013207} (\bibinfo {year} {2017})}\BibitemShut {NoStop}%
\bibitem [{\citenamefont {Mahajan}\ \emph {et~al.}(2024)\citenamefont
  {Mahajan}, \citenamefont {Sharma},\ and\ \citenamefont
  {Lingam}}]{mahajan2024}%
  \BibitemOpen
  \bibfield  {author} {\bibinfo {author} {\bibfnamefont {S.~M.}\ \bibnamefont
  {Mahajan}}, \bibinfo {author} {\bibfnamefont {P.}~\bibnamefont {Sharma}},\
  and\ \bibinfo {author} {\bibfnamefont {M.}~\bibnamefont {Lingam}},\
  }\bibfield  {title} {\bibinfo {title} {Hall mhd waves: A fundamental
  departure from their mhd counterparts},\ }\href
  {https://doi.org/10.1063/5.0227375} {\bibfield  {journal} {\bibinfo
  {journal} {Physics of Plasmas}\ }\textbf {\bibinfo {volume} {31}},\ \bibinfo
  {pages} {090701} (\bibinfo {year} {2024})}\BibitemShut {NoStop}%
\bibitem [{\citenamefont {Chandrasekhar}(1953)}]{chandrasekhar1953problems}%
  \BibitemOpen
  \bibfield  {author} {\bibinfo {author} {\bibfnamefont {S.}~\bibnamefont
  {Chandrasekhar}},\ }\bibfield  {title} {\bibinfo {title} {{Problems of
  stability in hydrodynamics and hydromagnetics (George Darwin Lecture)}},\
  }\href {https://doi.org/10.1093/mnras/113.6.667} {\bibfield  {journal}
  {\bibinfo  {journal} {Monthly Notices of the Royal Astronomical Society}\
  }\textbf {\bibinfo {volume} {113}},\ \bibinfo {pages} {667} (\bibinfo {year}
  {1953})}\BibitemShut {NoStop}%
\bibitem [{\citenamefont {Lehnert}(1954)}]{lehnert1954magnetohydrodynamic}%
  \BibitemOpen
  \bibfield  {author} {\bibinfo {author} {\bibfnamefont {B.}~\bibnamefont
  {Lehnert}},\ }\bibfield  {title} {\bibinfo {title} {{Magnetohydrodynamic
  Waves Under the Action of the Coriolis Force.}},\ }\href
  {https://doi.org/10.1086/145869} {\bibfield  {journal} {\bibinfo  {journal}
  {Astrophysical Journal}\ }\textbf {\bibinfo {volume} {119}},\ \bibinfo
  {pages} {647} (\bibinfo {year} {1954})}\BibitemShut {NoStop}%
\bibitem [{\citenamefont {Balbus}\ and\ \citenamefont
  {Terquem}(2001)}]{balbus2001}%
  \BibitemOpen
  \bibfield  {author} {\bibinfo {author} {\bibfnamefont {S.~A.}\ \bibnamefont
  {Balbus}}\ and\ \bibinfo {author} {\bibfnamefont {C.}~\bibnamefont
  {Terquem}},\ }\bibfield  {title} {\bibinfo {title} {Linear analysis of the
  hall effect in protostellar disks},\ }\href {https://doi.org/10.1086/320452}
  {\bibfield  {journal} {\bibinfo  {journal} {The Astrophysical Journal}\
  }\textbf {\bibinfo {volume} {552}},\ \bibinfo {pages} {235} (\bibinfo {year}
  {2001})}\BibitemShut {NoStop}%
\bibitem [{\citenamefont {Rax}\ \emph {et~al.}(2023)\citenamefont {Rax},
  \citenamefont {Gueroult},\ and\ \citenamefont {Fisch}}]{rax2023rotating}%
  \BibitemOpen
  \bibfield  {author} {\bibinfo {author} {\bibfnamefont {J.-M.}\ \bibnamefont
  {Rax}}, \bibinfo {author} {\bibfnamefont {R.}~\bibnamefont {Gueroult}},\ and\
  \bibinfo {author} {\bibfnamefont {N.}~\bibnamefont {Fisch}},\ }\bibfield
  {title} {\bibinfo {title} {Rotating alfvén waves in rotating plasmas},\
  }\href {https://doi.org/10.1017/S0022377823001368} {\bibfield  {journal}
  {\bibinfo  {journal} {Journal of Plasma Physics}\ }\textbf {\bibinfo {volume}
  {89}},\ \bibinfo {pages} {905890613} (\bibinfo {year} {2023})}\BibitemShut
  {NoStop}%
\bibitem [{\citenamefont {Turi}\ and\ \citenamefont
  {Misra}(2022)}]{turi2022magnetohydrodynamic}%
  \BibitemOpen
  \bibfield  {author} {\bibinfo {author} {\bibfnamefont {J.}~\bibnamefont
  {Turi}}\ and\ \bibinfo {author} {\bibfnamefont {A.~P.}\ \bibnamefont
  {Misra}},\ }\bibfield  {title} {\bibinfo {title} {Magnetohydrodynamic
  instabilities in a self-gravitating rotating cosmic plasma},\ }\href
  {https://doi.org/10.1088/1402-4896/ac9ca6} {\bibfield  {journal} {\bibinfo
  {journal} {Physica Scripta}\ }\textbf {\bibinfo {volume} {97}},\ \bibinfo
  {pages} {125603} (\bibinfo {year} {2022})}\BibitemShut {NoStop}%
\bibitem [{\citenamefont {Sharma}\ and\ \citenamefont
  {Patidar}(2021)}]{sharma2021modes}%
  \BibitemOpen
  \bibfield  {author} {\bibinfo {author} {\bibfnamefont {P.}~\bibnamefont
  {Sharma}}\ and\ \bibinfo {author} {\bibfnamefont {A.}~\bibnamefont
  {Patidar}},\ }\bibfield  {title} {\bibinfo {title} {{Modes of propagation and
  instabilities in finitely conducting neutrino-modified magnetohydrodynamic
  viscous plasma}},\ }\href {https://doi.org/10.1063/5.0044746} {\bibfield
  {journal} {\bibinfo  {journal} {Physics of Plasmas}\ }\textbf {\bibinfo
  {volume} {28}},\ \bibinfo {pages} {052116} (\bibinfo {year}
  {2021})}\BibitemShut {NoStop}%
\bibitem [{\citenamefont {Hager}\ \emph {et~al.}(2023)\citenamefont {Hager},
  \citenamefont {Khaled},\ and\ \citenamefont
  {Shukri}}]{hager2023magnetosonic}%
  \BibitemOpen
  \bibfield  {author} {\bibinfo {author} {\bibfnamefont {Y.~A.~A.}\
  \bibnamefont {Hager}}, \bibinfo {author} {\bibfnamefont {M.~A.~H.}\
  \bibnamefont {Khaled}},\ and\ \bibinfo {author} {\bibfnamefont {M.~A.}\
  \bibnamefont {Shukri}},\ }\bibfield  {title} {\bibinfo {title} {Magnetosonic
  waves propagation in a magnetorotating quantum plasma},\ }\href
  {https://doi.org/10.1103/PhysRevE.107.055202} {\bibfield  {journal} {\bibinfo
   {journal} {Phys. Rev. E}\ }\textbf {\bibinfo {volume} {107}},\ \bibinfo
  {pages} {055202} (\bibinfo {year} {2023})}\BibitemShut {NoStop}%
\bibitem [{\citenamefont {Turi}\ and\ \citenamefont
  {Misra}(2024)}]{turi2024nonlinear}%
  \BibitemOpen
  \bibfield  {author} {\bibinfo {author} {\bibfnamefont {J.}~\bibnamefont
  {Turi}}\ and\ \bibinfo {author} {\bibfnamefont {A.~P.}\ \bibnamefont
  {Misra}},\ }\bibfield  {title} {\bibinfo {title} {Nonlinear modulation of
  dispersive fast magnetosonic waves in an inhomogeneous rotating solar
  low-beta magnetoplasma},\ }\href {https://doi.org/10.1088/1402-4896/ad720a}
  {\bibfield  {journal} {\bibinfo  {journal} {Physica Scripta}\ }\textbf
  {\bibinfo {volume} {99}},\ \bibinfo {pages} {105602} (\bibinfo {year}
  {2024})}\BibitemShut {NoStop}%
\bibitem [{\citenamefont {{Ballester, J. L.}}\ \emph
  {et~al.}(2020)\citenamefont {{Ballester, J. L.}}, \citenamefont {{Soler,
  R.}}, \citenamefont {{Terradas, J.}},\ and\ \citenamefont {{Carbonell,
  M.}}}]{ballester2020}%
  \BibitemOpen
  \bibfield  {author} {\bibinfo {author} {\bibnamefont {{Ballester, J. L.}}},
  \bibinfo {author} {\bibnamefont {{Soler, R.}}}, \bibinfo {author}
  {\bibnamefont {{Terradas, J.}}},\ and\ \bibinfo {author} {\bibnamefont
  {{Carbonell, M.}}},\ }\bibfield  {title} {\bibinfo {title} {Nonlinear
  coupling of alfvén and slow magnetoacoustic waves in partially ionized solar
  plasmas},\ }\href {https://doi.org/10.1051/0004-6361/202038220} {\bibfield
  {journal} {\bibinfo  {journal} {A\&A}\ }\textbf {\bibinfo {volume} {641}},\
  \bibinfo {pages} {A48} (\bibinfo {year} {2020})}\BibitemShut {NoStop}%
\bibitem [{\citenamefont {Andrés}\ \emph {et~al.}(2017)\citenamefont
  {Andrés}, \citenamefont {Clark~di Leoni}, \citenamefont {Mininni},
  \citenamefont {Dmitruk}, \citenamefont {Sahraoui},\ and\ \citenamefont
  {Matthaeus}}]{andres2017}%
  \BibitemOpen
  \bibfield  {author} {\bibinfo {author} {\bibfnamefont {N.}~\bibnamefont
  {Andrés}}, \bibinfo {author} {\bibfnamefont {P.}~\bibnamefont {Clark~di
  Leoni}}, \bibinfo {author} {\bibfnamefont {P.~D.}\ \bibnamefont {Mininni}},
  \bibinfo {author} {\bibfnamefont {P.}~\bibnamefont {Dmitruk}}, \bibinfo
  {author} {\bibfnamefont {F.}~\bibnamefont {Sahraoui}},\ and\ \bibinfo
  {author} {\bibfnamefont {W.~H.}\ \bibnamefont {Matthaeus}},\ }\bibfield
  {title} {\bibinfo {title} {Interplay between alfvén and magnetosonic waves
  in compressible magnetohydrodynamics turbulence},\ }\href
  {https://doi.org/10.1063/1.4997990} {\bibfield  {journal} {\bibinfo
  {journal} {Physics of Plasmas}\ }\textbf {\bibinfo {volume} {24}},\ \bibinfo
  {pages} {102314} (\bibinfo {year} {2017})}\BibitemShut {NoStop}%
\bibitem [{\citenamefont {Dedner}\ \emph {et~al.}(2002)\citenamefont {Dedner},
  \citenamefont {Kemm}, \citenamefont {Kröner}, \citenamefont {Munz},
  \citenamefont {Schnitzer},\ and\ \citenamefont
  {Wesenberg}}]{dedner2002hyperbolic}%
  \BibitemOpen
  \bibfield  {author} {\bibinfo {author} {\bibfnamefont {A.}~\bibnamefont
  {Dedner}}, \bibinfo {author} {\bibfnamefont {F.}~\bibnamefont {Kemm}},
  \bibinfo {author} {\bibfnamefont {D.}~\bibnamefont {Kröner}}, \bibinfo
  {author} {\bibfnamefont {C.-D.}\ \bibnamefont {Munz}}, \bibinfo {author}
  {\bibfnamefont {T.}~\bibnamefont {Schnitzer}},\ and\ \bibinfo {author}
  {\bibfnamefont {M.}~\bibnamefont {Wesenberg}},\ }\bibfield  {title} {\bibinfo
  {title} {Hyperbolic divergence cleaning for the mhd equations},\ }\href
  {https://doi.org/https://doi.org/10.1006/jcph.2001.6961} {\bibfield
  {journal} {\bibinfo  {journal} {Journal of Computational Physics}\ }\textbf
  {\bibinfo {volume} {175}},\ \bibinfo {pages} {645} (\bibinfo {year}
  {2002})}\BibitemShut {NoStop}%
\bibitem [{\citenamefont {Bera}\ \emph {et~al.}(2022)\citenamefont {Bera},
  \citenamefont {Song},\ and\ \citenamefont {Srinivasan}}]{bera2022effect}%
  \BibitemOpen
  \bibfield  {author} {\bibinfo {author} {\bibfnamefont {R.~K.}\ \bibnamefont
  {Bera}}, \bibinfo {author} {\bibfnamefont {Y.}~\bibnamefont {Song}},\ and\
  \bibinfo {author} {\bibfnamefont {B.}~\bibnamefont {Srinivasan}},\ }\bibfield
   {title} {\bibinfo {title} {The effect of viscosity and resistivity on
  rayleigh–taylor instability induced mixing in magnetized
  high-energy-density plasmas},\ }\href
  {https://doi.org/10.1017/S0022377821001343} {\bibfield  {journal} {\bibinfo
  {journal} {Journal of Plasma Physics}\ }\textbf {\bibinfo {volume} {88}},\
  \bibinfo {pages} {905880209} (\bibinfo {year} {2022})}\BibitemShut {NoStop}%
\bibitem [{\citenamefont {Murawski}(1992)}]{murawski1992alfven}%
  \BibitemOpen
  \bibfield  {author} {\bibinfo {author} {\bibfnamefont {K.}~\bibnamefont
  {Murawski}},\ }\bibfield  {title} {\bibinfo {title} {Alfven-magnetosonic
  waves interaction},\ }\href@noop {} {\bibfield  {journal} {\bibinfo
  {journal} {Acta Physica Polonica A}\ }\textbf {\bibinfo {volume} {81}},\
  \bibinfo {pages} {335} (\bibinfo {year} {1992})}\BibitemShut {NoStop}%
\bibitem [{\citenamefont {Opat}(1990)}]{opat1990coriolis}%
  \BibitemOpen
  \bibfield  {author} {\bibinfo {author} {\bibfnamefont {G.~I.}\ \bibnamefont
  {Opat}},\ }\bibfield  {title} {\bibinfo {title} {{Coriolis and magnetic
  forces: The gyrocompass and magnetic compass as analogs}},\ }\href
  {https://doi.org/10.1119/1.16498} {\bibfield  {journal} {\bibinfo  {journal}
  {American Journal of Physics}\ }\textbf {\bibinfo {volume} {58}},\ \bibinfo
  {pages} {1173} (\bibinfo {year} {1990})}\BibitemShut {NoStop}%
\bibitem [{\citenamefont {Misra}\ \emph {et~al.}(2009)\citenamefont {Misra},
  \citenamefont {Ghosh},\ and\ \citenamefont {Shukla}}]{misra2009}%
  \BibitemOpen
  \bibfield  {author} {\bibinfo {author} {\bibfnamefont {A.~P.}\ \bibnamefont
  {Misra}}, \bibinfo {author} {\bibfnamefont {N.~K.}\ \bibnamefont {Ghosh}},\
  and\ \bibinfo {author} {\bibfnamefont {P.~K.}\ \bibnamefont {Shukla}},\
  }\bibfield  {title} {\bibinfo {title} {Evolution of alfvénic wave envelopes
  in spin-1/2 quantum hall-magnetohydrodynamic plasmas},\ }\href
  {https://doi.org/10.1063/1.3250987} {\bibfield  {journal} {\bibinfo
  {journal} {Physics of Plasmas}\ }\textbf {\bibinfo {volume} {16}},\ \bibinfo
  {pages} {102309} (\bibinfo {year} {2009})}\BibitemShut {NoStop}%
\bibitem [{\citenamefont {Chatterjee}\ and\ \citenamefont
  {Misra}(2018)}]{chatterjee2018}%
  \BibitemOpen
  \bibfield  {author} {\bibinfo {author} {\bibfnamefont {D.}~\bibnamefont
  {Chatterjee}}\ and\ \bibinfo {author} {\bibfnamefont {A.~P.}\ \bibnamefont
  {Misra}},\ }\bibfield  {title} {\bibinfo {title} {Modulation of kinetic
  alfvén waves in an intermediate low-beta magnetoplasma},\ }\href
  {https://doi.org/10.1063/1.5025895} {\bibfield  {journal} {\bibinfo
  {journal} {Physics of Plasmas}\ }\textbf {\bibinfo {volume} {25}},\ \bibinfo
  {pages} {052121} (\bibinfo {year} {2018})}\BibitemShut {NoStop}%
\bibitem [{\citenamefont {Ohtani}\ \emph {et~al.}(1989)\citenamefont {Ohtani},
  \citenamefont {Miura},\ and\ \citenamefont {Tamao}}]{ohtani1989}%
  \BibitemOpen
  \bibfield  {author} {\bibinfo {author} {\bibfnamefont {S.}~\bibnamefont
  {Ohtani}}, \bibinfo {author} {\bibfnamefont {A.}~\bibnamefont {Miura}},\ and\
  \bibinfo {author} {\bibfnamefont {T.}~\bibnamefont {Tamao}},\ }\bibfield
  {title} {\bibinfo {title} {Coupling between alfvén and slow magnetosonic
  waves in an inhomogeneous finite-beta plasma-i. coupled equations and
  physical mechanism},\ }\href
  {https://doi.org/https://doi.org/10.1016/0032-0633(89)90097-4} {\bibfield
  {journal} {\bibinfo  {journal} {Planetary and Space Science}\ }\textbf
  {\bibinfo {volume} {37}},\ \bibinfo {pages} {567} (\bibinfo {year}
  {1989})}\BibitemShut {NoStop}%
\end{thebibliography}%
\nopagebreak

\end{document}